\documentclass[aps,pre,amsmath,amssymb,lengthcheck,showpacs,superscriptaddress]{revtex4-1}
\usepackage{graphicx}
\usepackage{color}
\usepackage{subfigure}
\newcommand{\deri}[2]{{\displaystyle \frac{\partial #1 }{\partial #2 }}}

\newcommand{\derrri}[3]{{\displaystyle \frac{\partial^2 #1 }{\partial #2 \partial #3}}}

\begin{document}

\title{Cut-off nonlinearities in the low-temperature vibrations of glasses and crystals}

\author{Hideyuki Mizuno}
\email{hideyuki.mizuno@fukui.kyoto-u.ac.jp} 
\altaffiliation{Current address: Fukui Institute for Fundamental Chemistry, Kyoto University, Kyoto 606-8103, Japan}
\affiliation{Institut f\"{u}r Materialphysik im Weltraum, Deutsches Zentrum f\"{u}r Luft- und Raumfahrt (DLR), 51170 K\"{o}ln, Germany}

\author{Leonardo E.~Silbert} 
\affiliation{Department of Physics, Southern Illinois University Carbondale, Carbondale, IL 62901, USA}

\author{Matthias Sperl}
\affiliation{Institut f\"{u}r Materialphysik im Weltraum, Deutsches Zentrum f\"{u}r Luft- und Raumfahrt (DLR), 51170 K\"{o}ln, Germany}

\author{Stefano Mossa}
\affiliation{Univ. Grenoble Alpes, INAC-SPRAM, F-38000 Grenoble, France}
\affiliation{CNRS, INAC-SPRAM, F-38000 Grenoble, France}
\affiliation{CEA, INAC-SPRAM, F-38000 Grenoble, France}

\author{Jean-Louis Barrat}
\affiliation{Univ. Grenoble Alpes, LIPHY, F-38000 Grenoble, France}
\affiliation{CNRS, LIPHY, F-38000 Grenoble, France}
\affiliation{Institut Laue-Langevin - 6 rue Jules Horowitz, BP 156, 38042 Grenoble, France}

\date{\today}

\begin{abstract}
We present a computer simulation study of glassy and crystalline states using the standard Lennard-Jones interaction potential that is truncated at a finite cut-off distance, as is typical of many computer simulations.
We demonstrate that the discontinuity at the cut-off distance in the first derivative of the potential (corresponding to the interparticle force) leads to the appearance of cut-off nonlinearities.
These cut-off nonlinearities persist into the very-low-temperature regime thereby affecting low-temperature thermal vibrations, which leads to a breakdown of the harmonic approximation for many eigen modes, particularly for low-frequency vibrational modes.
Furthermore, while expansion nonlinearities which are due to higher order terms in the Taylor expansion of the interaction potential that are usually ignored at low temperatures and show up as the temperature increases, cut-off nonlinearities can become most significant at the lowest temperatures.
Anharmonic effects readily show up in the elastic moduli which not only depend on the eigen frequencies, but are crucially sensitive to the eigen vectors of the normal modes.
Whereas, those observables that rely mainly on static structural information or just the eigen frequencies, such as the vibrational density of states, total potential energy, and specific heat, show negligible dependence on the presence of the cut-off.
Similar aspects of nonlinear behavior have recently been reported in model granular materials, where the constituent particles interact through, finite-range, purely-repulsive, potentials.
These nonlinearities have been ascribed to the nature of the sudden cut-off at contact in the force-law, thus we demonstrate that cut-off nonlinearities emerge as a general feature of ordered and disordered solid state systems interacting through truncated potentials.
\end{abstract}


\maketitle

\section{Introduction}
Let us consider a Hamiltonian system, composed of $N$ point particles, in a 3-dimensional solid (crystalline or glassy) state.
We assume that the potential energy, $\Phi=\Phi(\{\vec{r}_i\})$, of the system is a function only of the positions of the particles, $\{ \vec{r}_1,\vec{r}_2,...,\vec{r}_N \}$.
When the amplitude of vibrational motions of the particles is small enough, the harmonic approximation provides a suitable description of the system~\cite{Ashcroft}, where the equations of motion are linearized
\begin{equation}
m_i \frac{d^2 \vec{u}_i}{dt^2} = -\sum_{j=1}^N \left(\derrri{\Phi}{\vec{r}_i}{\vec{r}_j}\right)^0 \cdot \vec{u}_j \quad (i=1,2,...,N). \label{harmonic}
\end{equation}
Here, $m_i$ is the mass of particle $i$, $\vec{u}_i(t) = \vec{r}_i(t)-\vec{r}_{i}^0$ is the displacement of particle $i$ from the potential minimum state ``$0$" where $\Phi=\Phi^0=\Phi(\{\vec{r}_i=\vec{r}_i^0\})$, and $()^0$ denotes the value at the minimum.
The solution of Eq.~(\ref{harmonic}) can be written as the superposition of $3(N-1)$ eigen modes ($3$ zero-frequency translational modes are removed)
\begin{equation}
\vec{u}_i(t) = \sum_{k=1}^{3N-3} A^k_0 \exp(j \omega^k t) \vec{e}^k_i \quad (i=1,2,...,N), \label{solution}
\end{equation}
where $A^k_0$ is the amplitude of mode $k$, and $j$ is the imaginary unit.
$\omega^k$ and $\vec{e}^k = [\vec{e}^k_1,\vec{e}^k_2,...,\vec{e}^k_N]$ are respectively the eigen frequency and the eigen vector of mode $k$, with $\vec{e}^k_{i}$ the polarization vector of particle $i$, which are obtained by solving the eigenvalue problem of the $3N \times 3N$ dynamical matrix, or Hessian, $\left({\partial^2 \Phi}/\partial\vec{r}_i \partial \vec{r}_j \right)^0$~\cite{Ashcroft}.
For thermal systems, e.g. crystals and glasses, in the low temperature regime such that thermal motions of the particles are constrained within their local energy minima, the harmonic approximation successfully describes both vibrational properties of the systems, and other thermodynamic observables, e.g. the specific heat and elastic constants.
The success of this picture relies on the notion that particle vibrations are analogous to a ball-spring model whereby interacting particles remain in `contact' in the sense that particles always feel the interactions of their neighbors.

In contrast, recent studies of purely-repulsive, short-range interaction potentials as model systems of particulate matter have revealed a strong nonlinear or anharmonic feature in their vibrational modes~\cite{Schreck_2011,Bertrand_2014,Schreck_2014,Ikeda_2013,Goodrich_2014,Goodrich2_2014}.
This nonlinear behavior has been ascribed to the nature of the sudden cut-off at contact in the force-law between non-cohesive granular particles.
Indeed, given the one-sided nature of such potentials and the fact that their second derivatives are non-analytic at the contact distance~\cite{OHern_2003}, once a contact altering event (opening or closing) occurs, the harmonic approximation, Eq.~(\ref{harmonic}), cannot capture this energy variation, and nonlinearities emerge~\cite{Schreck_2011,Bertrand_2014,Schreck_2014,Ikeda_2013,Goodrich_2014,Goodrich2_2014}.
We denote these aspects of the interaction law as ``cut-off nonlinearities'' and emphasize that these are quite distinct from the usual ``expansion nonlinearities'' which arise from higher order terms in the Taylor expansion of the potential energy, $\Phi=\Phi(\{\vec{r}_i\})$, around the energy minimum~\cite{Ikeda_2013,Hentschel_2011}.

In the present contribution, we show that cut-off nonlinearities are not specific to granular matter, but appear as a general feature in any solid state system in which the interaction is cut off at a finite distance, as is common practice in computer simulations.
While enforcing the truncation of quasi-long-ranged interaction potentials, e.g., attractive Lennard-Jones potentials, is largely a numerical issue to improve computational efficiency, from a mathematical perspective the appearance of cut-off nonlinearities that results from this truncation share the same mechanism with those in particulate, jammed systems.
Therefore, it seems pertinent to ask whether such cut-off nonlinearities actually affect structural and mechanical properties of solid state systems.
We will show that, although simple structural properties are not influenced by these nonlinearities, they become apparent in normal modes analyses.
In particular, such cut-off nonlinearities can strongly affect mechanical properties that depend explicitly on the vibrational modes, specifically, the eigen vectors and thence the particle polarization vectors.
Oddly, this effect can be enhanced by lowering the temperature.

We study the Lennard-Jones (LJ) pair potential $\phi_{{LJ}}(r)$ (our results for the LJ potential can be extended to more practical potentials, and should be quite general)
\begin{equation}
\begin{aligned}
\phi_{{LJ}}(r) &= 4\epsilon \left[ \left(\frac{\sigma}{r} \right)^{12} - \left(\frac{\sigma}{r} \right)^6 \right], \\
\phi'_{{LJ}}(r) &= -\frac{24 \epsilon}{r} \left[ 2\left(\frac{\sigma}{r} \right)^{12} - \left(\frac{\sigma}{r} \right)^6 \right],
\end{aligned}
\end{equation}
where $'$ denotes the derivative with respect to distance $r$.
The values of $\epsilon$ and $\sigma$ characterize the energy and length scales of the interaction, respectively.
The standard protocol in numerical simulations is to truncate the potential at some cut-off distance $r=r_c$ \cite{Allen1986}
\begin{equation}
\begin{aligned}
\phi_{TLJ}(r) &= \phi_{{LJ}}(r) H(r_c-r), \\
\phi'_{TLJ}(r) &= \phi'_{{LJ}}(r) H(r_c-r) - \phi_{{LJ}}(r) \delta(r-r_c),
\end{aligned} \label{tpotential}
\end{equation}
where $H(x)$ is the Heaviside function ($H(x)=1$ for $x \ge 0$, $H(x)=0$ otherwise), and $\delta(x)$ is the impulsive function.
This truncation leads to a discontinuity at $r=r_c$ in the potential $\phi_{TLJ}(r)$, and also an impulsive term, $\sim \delta(r-r_c)$, in its first derivative $\phi'_{TLJ}(r)$, i.e. the interparticle force.
To prevent the discontinuity in the potential and impulsive force, it is standard practice to use the ``shifted potential" $\phi_{SP}(r)$~\cite{Allen1986,Xu_2012};
\begin{equation}
\begin{aligned}
\phi_{SP}(r) &= \left[ \phi_{{LJ}}(r) - \phi_{{LJ}}(r_c) \right] H(r_c-r), \\
\phi'_{SP}(r) &= \phi'_{{LJ}}(r) H(r_c-r).
\end{aligned} \label{spotential}
\end{equation}
The function $\phi_{SP}(r)$ is now continuous at $r=r_c$ and has no impulsive term in its first derivative, but still has discontinuities at $r=r_c$ in its derivatives, including the force term.
To smooth the potential further, e.g. make the first derivate continuous, we can employ the ``shifted-force" potential $\phi_{SF}(r)$~\cite{Allen1986,Toxvaerd_2011};
\begin{equation}
\begin{aligned}
\phi_{SF}(r) &= \left[ \phi_{{LJ}}(r) - \phi_{{LJ}}(r_c)- (r-r_c)\phi'_{{LJ}}(r_c) \right] H(r_c-r), \\
\phi'_{SF}(r) &= \left[ \phi'_{{LJ}}(r)-\phi'_{{LJ}}(r_c) \right] H(r_c-r).
\end{aligned} \label{sforce}
\end{equation}
We can also smooth arbitrarily high $n$th-order ($n\ge 2$) derivatives by adding terms, $\sim (r-r_c)^n$, in the same way as in $\phi_{SF}(r)$.
Another option is to interpolate the potential around the cut-off distance $r=r_c$ by using polynomials (see e.g. Ref.~\cite{Voigtmann_2009}).

If the second derivative of the truncated potential is non-analytic at $r=r_c$, as it is the case for all of $\phi_{TLJ}(r), \phi_{SP}(r), \phi_{SF}(r)$, and if some pairs of particles pass through $r=r_c$, the dynamical matrix and Eq.~(\ref{harmonic}) cannot capture the energy change due to this event, and as a result, cut-off nonlinearities emerge.
In the following, we will show that the cut-off discontinuity in the interparticle force (the first derivative of the potential) enhances the cut-off nonlinearities, and has non-negligible effects on the low-temperature thermal vibrations, even though the discontinuity is small.
Specifically, when we use the shifted-potential $\phi_{SP}(r)$, where the interparticle force is discontinuous at $r=r_c$, the harmonic approximation breaks down even at very low temperatures for many vibrational eigen modes, particularly those that lie at lower frequencies.
However, when the shifted-force potential $\phi_{SF}(r)$ is employed to make the interparticle force continuous, then the cut-off nonlinearities are significantly suppressed, and the harmonic approximation becomes applicable again.
We also study the effects of cut-off nonlinearities on several structural and thermodynamic observables.
In particular, we will show that the cut-off nonlinearities have a negligible effect on those properties that do not depend on the details of the eigen vectors of the normal mode decomposition.
These include the radial distribution function (RDF), vibrational density of states (vDOS), total potential energy, and specific heat.
Whereas, mechanical properties such as the elastic constants, which depend crucially on particle polarization vectors, are much more strongly affected.

The paper is organized as follows.
In Sec.~\ref{sec.system} we describe in detail the numerical systems studied.
Section~\ref{results} contains a comprehensive presentation of our results.
We first discuss the results of the static structure and vibrational states in the harmonic limit $T=0$ in Sec.~\ref{seczerot}.
We next study the effect of cut-off nonlinearities on thermal vibrations at finite temperatures $T>0$ in Sec.~\ref{seccutoff}, and their effects on several physical quantities in Sec.~\ref{effectofcutoff}.
We summarize our results in Sec~\ref{summary}, and give our conclusion in Sec.~\ref{conclusion}.

\begin{table}[t]
\centering
\renewcommand{\arraystretch}{1.1}
\begin{tabular}{ccccc}
\hline
\hline
Potential & Force at $r=r_c$    & System & $\Phi^0/N$ \\
\hline
 $1$. \ $\phi_{SP}(r)$ with $r_{c}=2.5$ & Discontinuous & Glass$1$   & $-6.85281$ \\
                                      &                 & Crystal$1$ & $-7.36841$ \\
\hline
 $2$. \ $\phi_{SF}(r)$ with $r_{c}=2.5$ & Continuous    & Glass$2$   & $-6.07778$ \\
                                      &                 & Crystal$2$ & $-6.60853$ \\
\hline
 $3$. \ $\phi_{SP}(r)$ with $r_{c}=3.0$ & Discontinuous & Glass$3$   & $-7.29178$ \\
                                      &                 & Crystal$3$ & $-7.81922$ \\
\hline
\hline
\end{tabular}
\caption{
The potentials and the systems investigated in this work.
We consider three types of truncated LJ potentials, $1$, $2$, and $3$, and two types of configurations, glass and face centered cubic ($FCC$) crystal, therefore a total of $6=3\times 2$ systems, which are listed in the table.
We also present the value of the minimum potential energy per particle, $\Phi^0/N$, for each system.
The detailed descriptions of the potentials and the systems are found in the main text.
See also Fig.~\ref{ljpotential}.
}
\label{tsystem}
\end{table}

\begin{figure}[t]
\centering
\includegraphics[width=0.425\textwidth]{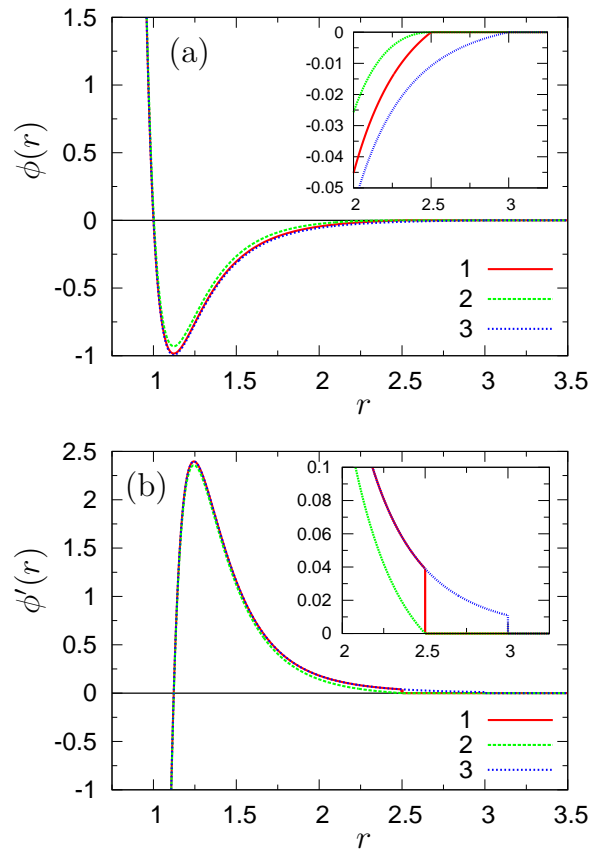}
\vspace*{0mm}
\caption{
(a) Lennard-Jones potential $\phi(r)$ and (b) its first derivative $\phi'(r)$.
In the present study, we use three types of truncated potentials; 1: $\phi_{SP}(r)$ with $r_c=2.5$, 2: $\phi_{SF}(r)$ with $r_c=2.5$, and 3: $\phi_{SP}(r)$ with $r_c=3.0$, as listed in Table~\ref{tsystem} (see also Eqs.~(\ref{spotential}) for $\phi_{SP}(r)$ and (\ref{sforce}) for $\phi_{SF}(r)$).
The red solid, green dashed, and blue dotted lines show the potentials 1, 2, and 3, respectively.
The insets show close-ups around the cut-off distance $r=r_c$.
}
\label{ljpotential}
\end{figure}

\section{Systems description} \label{sec.system}

\subsection{Lennard-Jones potentials}
We consider three types of truncated LJ potentials, as listed in Table~\ref{tsystem} and plotted in Fig.~\ref{ljpotential}.
Previous numerical works usually made the potential $\phi(r)$ continuous (shifted) at the cut-off distance $r=r_c$~\cite{Allen1986,Xu_2012,Toxvaerd_2011,Smit_1992,Shi_2001,Ahmed_2010}, in order to avoid the impulsive force, $\sim \delta(r-r_c)$.
Then, each work optionally took care of discontinuities in the derivatives of the potential.
Even if the potential $\phi(r)$ was not made continuous (not shifted) at $r=r_c$, the impulsive force was usually ignored, although the interparticle force and $\phi'(r)$ become inconsistent in this case.
Therefore, the shifted potential $\phi_{SP}(r)$ (see Eq.~(\ref{spotential})) is considered to be the most practical to implement and is regarded as the standard potential to use in simulation studies~\cite{Allen1986,Xu_2012,Toxvaerd_2011,Smit_1992,Shi_2001,Ahmed_2010}.
Our focus in this study is then the discontinuity in the first derivative of the interaction potential (interparticle force), which in turn has a non-negligible effect on the low-temperature vibrational motions of particles.

{\bf Potential 1}: As the first potential ``$1$'' we use the shifted potential, $\phi_{SP}(r)$, with $r_{c}=2.5$.
This potential has a discontinuity in its first derivative at $r=r_c=2.5$.
The value of the cut-off distance, $r_c=2.5$, has been employed in many previous simulation works~\cite{Allen1986,Xu_2012,Toxvaerd_2011}.
Note that the value of original potential $\phi_{LJ}(r)$ at $r_c=2.5$ is $1.6\%$ of the energy scale $\epsilon$.
Our previous simulations of LJ glasses also used this potential $1$ (with $r_c=2.5$) to study acoustic excitations \cite{monaco2_2009} and spatial distributions of local elastic moduli \cite{Mizuno_2013}.

{\bf Potential 2}: In order to consider the continuous interparticle force, we use, as the second potential~``$2$", the shifted-force potential, $\phi_{SF}(r)$, with the same cut-off distance $r_{c}=2.5$ (see Eq.~(\ref{sforce})).

{\bf Potential 3}: For our third potential~``$3$", we consider a longer cut-off distance $r_c=3.0$ in the shifted potential $\phi_{SP}(r)$.
The interparticle force has a discontinuity at $r=r_c$, but it is reduced by the longer $r_c=3.0$, compared to that of potential $1$ with $r_c=2.5$, as we can visually recognize in the inset of Fig.~\ref{ljpotential}(b).

By comparing the results from those three potentials, we study the impact of the discontinuity in the interparticle force at $r=r_c$, on cut-off nonlinearities and finite-temperature thermal vibrations.

\subsection{Systems preparation}
For each LJ potential described above, we prepared two types of configurations, an amorphous glass and a face centered cubic ($FCC$) crystal, under periodic boundary conditions in the $x$, $y$, and $z$ directions.
The system is mono-disperse, composed of $N=4,000$ identical particles with mass $m$, diameter $\sigma$, and interparticle potential energy $\epsilon$.
Throughout this paper, we present the values of quantities in units of $\sigma$ (length), $\epsilon$ (energy), and $m$ (mass).
The temperature $T$ and frequency $\omega$ are expressed in units of $\epsilon/k_B$ ($k_B$ is Boltzmann's constant) and $\tau^{-1}=(m\sigma^2/\epsilon)^{-1/2}$, respectively.
We fixed the number density at $\hat{\rho}=N/V=1.015$, where the system length is $L=V^{1/3}=15.8$, and the lattice constant of the $FCC$ crystal is $a=L/10=1.58$.
The melting temperature $T_m$ and the glass transition temperature $T_g$ of this mono-disperse system have been reported as $T_m \simeq 1.0$ and $T_g \simeq 0.4$~\cite{robles_2003}.

We first considered potential $1$, to prepare ``glass $1$" and ``crystal $1$" at a temperature $T=10^{-3}$, well below both $T_g$ and $T_m$.
For the preparation of glass $1$, we equilibrated the system at $T=2.0$ in the normal liquid phase, and quenched it down to $T=10^{-3}$ with an extremely fast rate $dT/dt=4 \times 10^2$, followed by an equilibration run at $T=10^{-3}$, as described in Ref.~\cite{monaco2_2009}.
After preparing the glass and crystal with potential $1$, we switched the potential from $1$ to $2$ and $3$, and equilibrated the systems again at the same temperature $T=10^{-3}$.
For the glassy cases, we relaxed the systems for sufficiently long times to eliminate any aging effects in these three cases.
We note that all systems remain very close to the same energy minimum during the entire trajectory.
After the systems were equilibrated at $T=10^{-3}$, we quenched them using the steepest descent method, from $T=10^{-3}$ to $0$, i.e. into the nearest energy minimum $0$ (inherent structure).

\begin{figure}[t]
\centering
\includegraphics[width=0.425\textwidth]{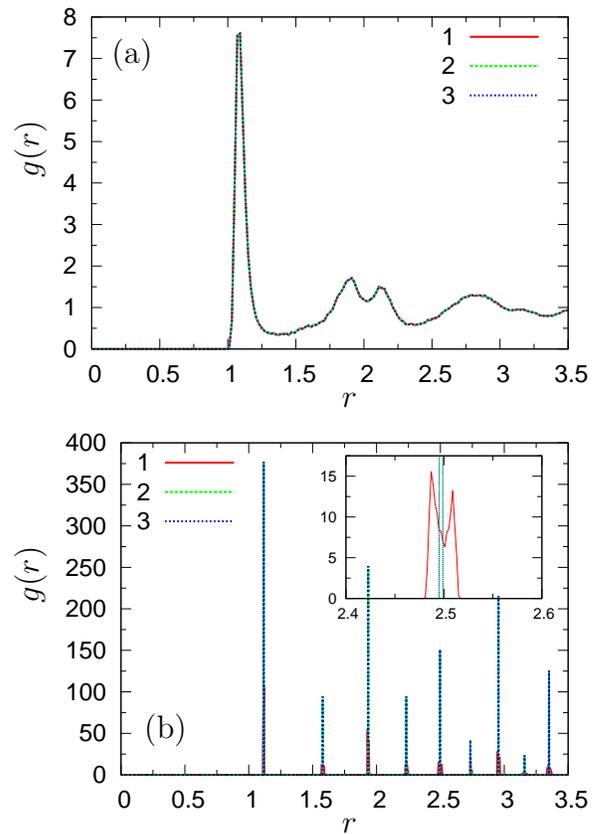}
\vspace*{0mm}
\caption{
The RDF $g(r)$ in the energy minimum state ($T=0$) for (a) glasses and (b) crystals.
The values for three types of truncated potentials, $1$ (red solid line), $2$ (green dashed line), and $3$ (blue dotted line), are presented.
The inset to (b) is a close-up around $r=2.5$.
}
\label{gr0}
\end{figure}

The values of the minimum potential energy per particle, $\Phi^0/N$, presented in Table~\ref{tsystem}, are different for the three potentials, with a difference of about $15\%$ between the potentials $2$ and $3$, for both glass and crystal.
Since different cut-offs give different Hamiltonians, system properties generally depend on the cut-off nature.
In fact, as it has been reported in Refs.~\cite{Smit_1992,Shi_2001,Ahmed_2010}, thermodynamic properties, including the melting point $T_m$ (and possibly the glass transition point $T_g$), strongly depend on the cut-off treatment.
However, as we will see in the RDF $g(r)$ of Fig.~\ref{gr0} in Sec.~\ref{secstatic}, the three truncated potentials produced practically identical configurations (glass or crystal).
This allows us to focus entirely on the role of the cut-off nonlinearities {\em in the same static structures}.
Note that we switched to potentials $2$ and $3$ only after the system was prepared with potential $1$; the potentials $2$ and $3$ were \textit{not} used from the initial stage where we generated liquid configuration at $T=2.0$.
The reason why we employed this particular preparation procedure was to minimize any possible differences in the configurations from which we started to compute observables.

Finally, each system was heated from the energy minimum state ($T=0$).
We study the low temperature regime, ranging from $T=10^{-4}$ to $10^{-2}$ for glasses, and from $T=10^{-4}$ to $10^{-1}$ for crystals.
This temperature range is well below (one order of magnitude lower than) the glass transition $T_g \simeq 0.4$ and the melting $T_m \simeq 1.0$ temperatures.
At each $T$, we carried out an equilibration followed by a production run, using $NVT$ molecular-dynamics (MD) simulation.
Here, we set the time step to $\delta t = 10^{-2}$ for $T<10^{-3}$, and $\delta t = 5\times 10 ^{-3}$ for $T \ge 10^{-3}$.
During the simulations, we observed no particle rearrangements, indicating that particles vibrate around the same energy minimum state at all studied temperatures, i.e. each system remains in its original inherent structure.
All MD simulations were performed using LAMMPS~\cite{Plimpton_1995,lammps}.

\section{Results} \label{results}

\subsection{Static structure and vibrational states in the harmonic limit, $T=0$} \label{seczerot}
Before studying thermal vibrations at finite temperatures $T>0$, we first look at the static structure and the vibrational states in the energy minimum state $T=0$ (the harmonic limit).
We will see that the three truncated LJ potentials produce practically identical static structures and vibrational states at $T=0$.

\subsubsection{Static structure} \label{secstatic}
Figure~\ref{gr0} presents the RDF, $g(r)$, in the energy minimum state $0$ ($T=0$).
As in Fig.~\ref{gr0}(a), the three potentials show no differences in $g(r)$ for the glasses.
In the crystals, we observe slight differences between potential $1$ and the other potentials $2, 3$, as highlighted in the inset of Fig.~\ref{gr0}(b).
While crystals $2$ and $3$ have a single peak around each lattice distance, double peaks are apparent in crystal $1$.
In the $FCC$ lattice structure with lattice constant $a=1.58$, all the particles have their nearest neighbors located at a distance $r=\left(\sqrt{10}/2 \right) \times a \simeq 2.5$.
In this situation, if the interparticle force has a discontinuity at $r=r_c = 2.5$, particles cannot be stabilized at the \textit{exact} lattice positions, rather they are slightly displaced due to the discontinuous force, leading to the double peaks in $g(r)$ for crystal $1$.
This effect also exists in glasses $1$ and $3$, i.e. some pairs of particles located at the cut-off distance $r=r_c$ are not stabilized precisely at $r=r_c$.
However, this feature is not noticeable in the glass due to the smoothing effect of the amorphous structure.

Although the locations of particles in crystal $1$ do not lie precisely at the lattice positions, they are approximately located at these lattice points, with deviations of only $1\%$, hence they show the same $FCC$ structure as crystals $2, 3$.
Therefore, we conclude that the three truncated potentials produce essentially identical static structures in the harmonic limit $T=0$.

\begin{figure}[t]
\centering
\includegraphics[width=0.425\textwidth]{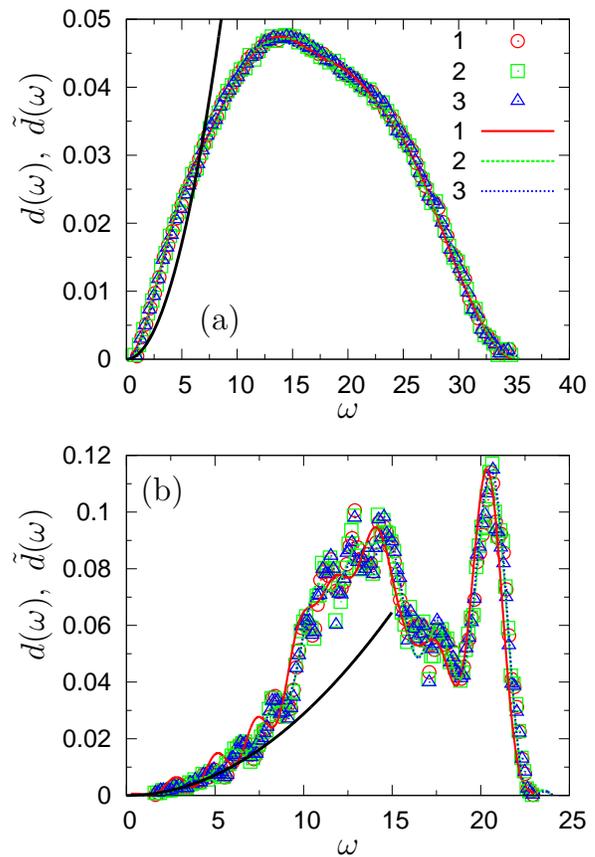}
\vspace*{0mm}
\caption{
The vDOS, $d(\omega)$ and $\tilde{d}(\omega)$, for (a) glasses and (b) crystals.
The symbols represent the vDOS $d(\omega)$ obtained from the dynamical matrix (see Eq.~(\ref{vdoseq})), for potentials, $1$ (red circle), $2$ (green square), and $3$ (blue triangle).
The lines show the Fourier transform, $\tilde{d}(\omega)$, of the velocity auto-correlation function, $d(t)$, at $T=10^{-3}$ (see Eq.~(\ref{vafeq})), for potentials, $1$ (red solid line), $2$ (green dashed line), and $3$ (blue dotted line).
The black solid line is the Debye prediction, which is calculated from the elastic moduli of potential $2$.
Note that the Debye prediction is almost unchanged for different potentials.  
}
\label{dos}
\end{figure}

\begin{figure}[t]
\centering
\includegraphics[width=0.425\textwidth]{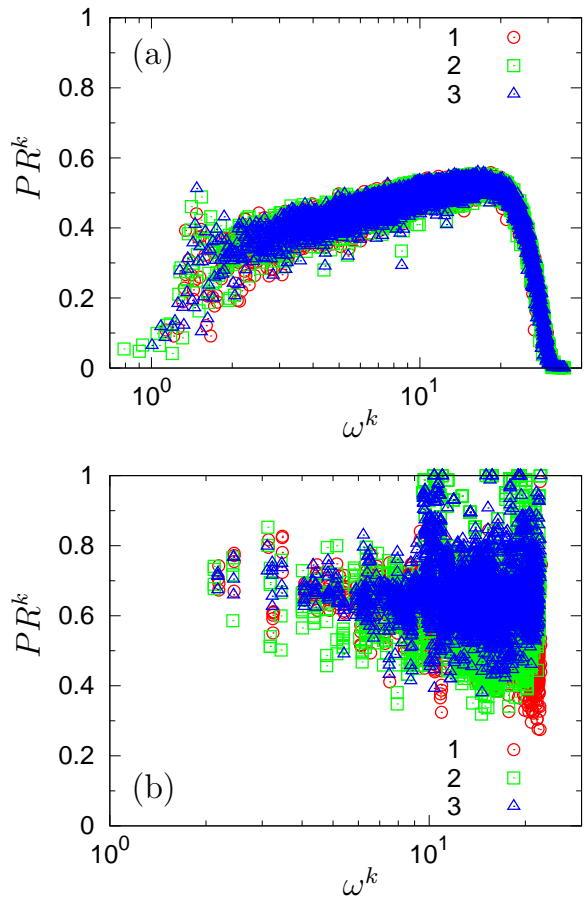}
\vspace*{0mm}
\caption{ 
Participation ratio $PR^k$ versus eigen frequency $\omega^k$ for (a) glasses and (b) crystals.
The values for the three potentials, $1$ (red circle), $2$ (green square), and $3$ (blue triangle), are presented.
}
\label{pr}
\end{figure}

\begin{figure}[t]
\centering
\includegraphics[width=0.425\textwidth]{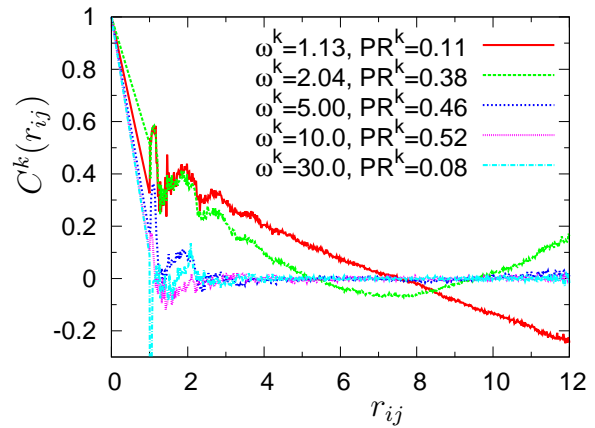}
\vspace*{0mm}
\caption{
The spatial correlation function $C^k(r_{ij})$ of eigen vector $\vec{e}^k_i(\vec{r}_i)$ for several different eigen modes in glass $1$.
Note that the mode with $\omega^k=1.13$ and $PR^k=0.11$ is a low frequency quasi-localized mode, while the one with $\omega^k=30.0$ and $PR^k=0.08$ is a high frequency localized mode~\cite{mazzacurati_1996,Schober_2004,Silbert_2009,Xu_2010}.
}
\label{length}
\end{figure}

\subsubsection{Vibrational states} \label{secharmonic} 
As we explained in Eqs.~(\ref{harmonic}) and~(\ref{solution}), harmonic vibrational motions of particles are completely described in terms of the $3N-3$ eigen frequencies $\omega^k$ and eigen vectors $\vec{e}^k=[\vec{e}^k_1,\vec{e}^k_2,...,\vec{e}^k_N]$ ($k=1,2,...,3N-3$)~\cite{Ashcroft}.
We diagonalized the Hessian ($3N \times 3N$ dynamical matrix), $({\partial^2 \Phi}/\partial\vec{r}_i \partial \vec{r}_j)^0$, to obtain the eigen values $\lambda^k$ and eigen vectors $\vec{e}^k$.
The eigen frequencies are given as $\omega^k = \sqrt{\lambda^k}$, and the eigen vectors are normalized such that $\vec{e}^k \cdot \vec{e}^l = \sum_{i=1}^N (\vec{e}_i^k \cdot \vec{e}_i^l)= \delta_{kl}$, where $\delta_{kl}$ is the Kronecker delta function.
Note that if the system is stable in the energy minimum state $0$, all the eigen values $\lambda^k$ are positive, and the eigen frequencies $\omega^k$ are real numbers.
In this study, we used the ARPACK software~\cite{arpack} to realize the diagonalization of the $3N \times 3N$ matrix.

Figure~\ref{dos} shows the vDOS, $d(\omega)$, obtained from $3N-3$ values of $\omega^k$;
\begin{equation}
d(\omega) = \frac{1}{3(N-1)} \sum_{k=1}^{3N-3} \delta(\omega-\omega^k). \label{vdoseq}
\end{equation}
In Fig.~\ref{pr}, we plot the participation ratio, $PR^k$, for each mode $k$, as a function of $\omega^k$;
\begin{equation}
PR^k = \frac{1}{N} \left[ \sum_{i=1}^N (\vec{e}_i^k \cdot \vec{e}_i^k)^2 \right]^{-1},
\end{equation}
which measures the extent of vibrational localization in mode $k$~\cite{mazzacurati_1996,Schober_2004,Silbert_2009,Xu_2010}.
The vDOS $d(\omega)$ shows the typical shapes for glasses and crystals.
{\bf Crystals:} (i) at intermediate-high frequencies $d(\omega)$ has two branches, corresponding to transverse and longitudinal acoustic phonons, and (ii) at low frequency, Debye scaling, $d(\omega) \sim \omega^{2}$, is observed (black solid line in Fig.~\ref{dos}(b))~\cite{Ashcroft,Mizuno2_2013}.
{\bf Glasses:} (i) smooth variations and broadened distributions are observed, and (ii) the low frequency portion of $d(\omega)$ shows an enhancement of modes over the Debye prediction~\cite{monaco2_2009,Mizuno2_2013,shintani_2008}.
In addition, from $PR^k$, we recognize localization of the low and high $\omega^k$ modes in the glasses, which are a characteristic feature of various amorphous~\cite{mazzacurati_1996,Schober_2004} and jammed~\cite{Silbert_2009,Xu_2010} matters, whereas all the modes in the crystals are extended phonon modes.

In Figs.~\ref{dos} and~\ref{pr}, there does not appear to be any noticeable differences among the three potentials.
This is further validated by the results shown in Fig.~\ref{fmom} (see lines) and related discussion.
Therefore, the three truncated potentials produce practically identical vibrational states in the harmonic limit $T=0$.
We note, however, that small detailed differences are observed in $d(\omega)$ and $PR^k$; for instance, glass $2$ with continuous interparticle force contains a number of lower frequency modes than those in glass $1$ (and $3$) with the discontinuous force (the lowest eigen frequency of $\omega^k=0.79$ in glass $2$ is lower than that of $\omega^k=1.13$ in glass $1$).
These small differences are picked up by the (non-affine) elastic moduli which shows differences between the three potentials, as indicated in Fig.~\ref{nonaffine} (see lines).

We close this section on the $T=0$ properties with a brief discussion of the low frequency quasi-localized modes in glassy configurations, which will be useful in the next section (Sec.~\ref{secvibamp}).
We have studied the spatial correlation function $C^k(r_{ij})$ of the eigen vectors $\vec{e}^k_i(\vec{r}_i)$ for each mode $k$;
\begin{equation}
C^k(r_{ij}) = \left< \vec{e}^k_i (\vec{r}_i) \cdot \vec{e}^k_j(\vec{r}_j) \right>_{\left<ij \right>},
\end{equation}
where $\vec{e}^k_i(\vec{r}_i)$ is a function of the position $\vec{r}_i$, $r_{ij}= |\vec{r}_i-\vec{r}_j|$, and $\left< \right>_{\left<ij \right>}$ denotes the average over all the pairs of particles $\left<ij \right>$.
Figure~\ref{length} presents $C^k(r_{ij})$ for several different eigen modes $k$ in glass $1$.
It is observed that the lowest $\omega^k=1.13$ mode has the low value of $PR^k=0.11$, but its $C^k(r_{ij})$ shows a longer spatial correlation compared to other modes with higher values of $PR^k$.
Note that the negative correlation comes from the transverse nature of this mode.
In the localized modes with low $PR^k$, vibrational motions of particles are confined to a few small regions.
Therefore one may expect that the spatial correlation between vibrational motions of particles is rather short within those confined regions.
However, Fig.~\ref{length} confirms that the lowest $\omega^k=1.13$ mode actually shows an extended character with the long spatial correlation.
Thus, the low $\omega^k$ mode is ``quasi-localized"~\cite{mazzacurati_1996,Schober_2004,Silbert_2009,Xu_2010}, which is distinct from ``true" localization without any extended nature, observed in the high frequency modes.
In Fig.~\ref{length}, the high frequency localized mode ($\omega^k=30.0$ and $PR^k=0.08$) shows only a short spatial correlation, i.e. $C^k(r_{ij})$ decays within $r_{ij} \sim 1$ (order of the particle diameter).
Similar observations have been found in athermal jammed solids~\cite{Silbert_2009,Xu_2010}.

\subsection{Cut-off nonlinearities in thermal vibrations at finite temperature $T>0$} \label{seccutoff} 
We now turn our attention to thermal vibrations at finite temperatures $T>0$.
We focus on each eigen mode $k$ and monitor vibrational motions of particles along the mode $k$.
Specifically, we measured the vibrational amplitude, $A^k(t)$, and the energy landscape, $\Delta \Phi (A^k)$, for each mode $k$, which will be presented in this section.
When a pair of particles passes through the cut-off distance $r=r_c$, the cut-off nonlinearities affect the vibrational motions of particles.
We will show that the discontinuity in the interparticle force at $r=r_c$ enhances the cut-off nonlinearities, which causes a breakdown of the harmonic approximation for many eigen modes $k$.
More precisely, the number of particles pairs that are involved in mode $k$ and experience the force discontinuity determines the strength of the cut-off nonlinearities.

\begin{figure*}[t]
\centering
\includegraphics[width=0.9\textwidth]{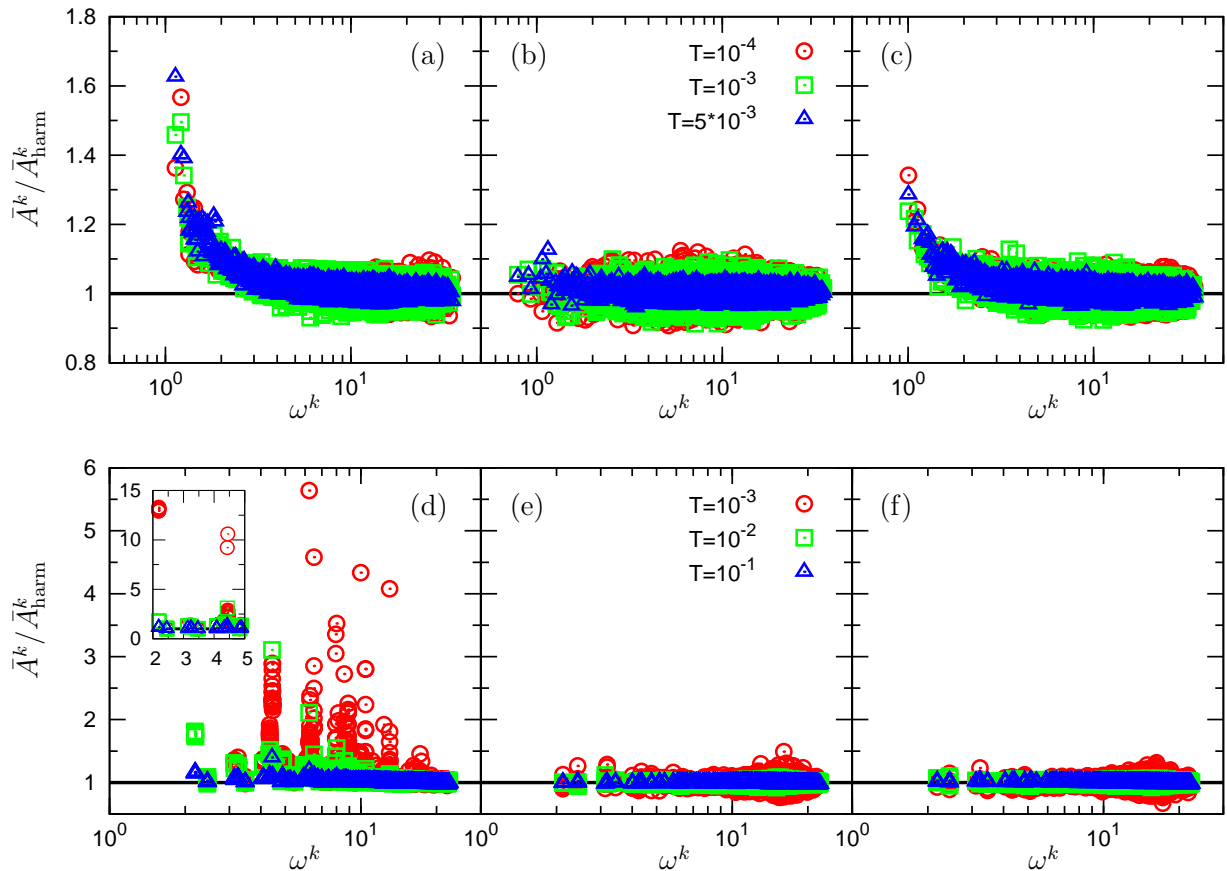}
\vspace*{0mm}
\caption{
Amplitude $\bar{A}^k$ versus eigen frequency $\omega^k$: (a) glass $1$, (b) glass $2$, (c) glass $3$, (d) crystal $1$, (e) crystal $2$, and (f) crystal $3$.
The value of $\bar{A}^k$ is normalized by the harmonic expectation, $\bar{A}^k_\text{harm}=\sqrt{T}/\omega^k$, such that harmonic behavior of mode $k$ corresponds to $\bar{A}^k/\bar{A}^k_\text{harm} = 1$ (Eq.~(\ref{requirement})), which is indicated by the horizontal solid line.
The inset to (d) shows the large values of the amplitude in crystal $1$.
Different symbols represent three temperatures as indicated in the key of the middle panel.
}
\label{amplitude}
\end{figure*}

\begin{figure}[t]
\centering
\includegraphics[width=0.425\textwidth]{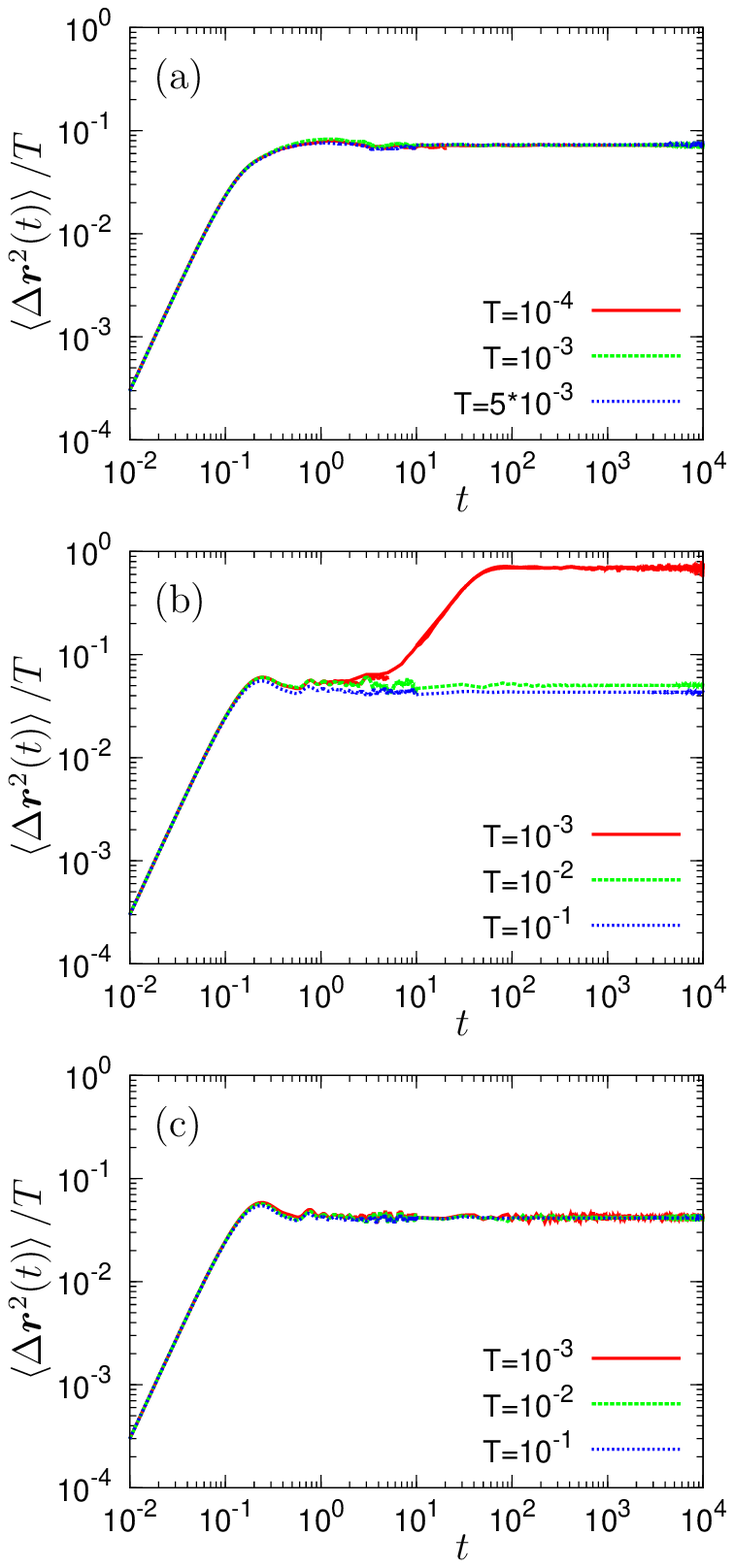}
\vspace*{0mm}
\caption{
Mean square displacement of particles, $\left< \Delta \vec{r}^2(t) \right>$ (Eq.~(\ref{emsd})), is plotted as a function of time $t$, for (a) glass $1$, (b) crystal $1$, and (c) crystal $2$.
The value of $\left< \Delta \vec{r}^2(t) \right>$ is normalized by the temperature $T$.
Different lines represent three temperatures as indicated in the key.
In the harmonic regime values of $\left< \Delta \vec{r}^2(t) \right>/T$ for different $T$ collapse onto a single curve.
}
\label{fmsd}
\end{figure}

\subsubsection{Vibrational amplitude along eigen mode $k$} \label{secvibamp}
We have extracted the vibrational amplitude, $A^k(t)$, along the eigen mode $k$ from an MD trajectory in the (NVT)-ensemble, $\{\vec{r}_1(t), \vec{r}_2(t), ..., \vec{r}_N(t) \}$, at each temperature $T$;
\begin{equation}
A^{k}(t) = \sum_{i=1}^N \vec{u}_i(t) \cdot \vec{e}^k_i \quad (k=1,2,...,3N-3), \label{ampleq}
\end{equation}
where $\vec{u}_i(t)=\vec{r}_i(t)-\vec{r}_i^0$ is the displacement of particle $i$ from its energy minimum position $\vec{r}_i^0$.
We note that the MD trajectory is then described as the superposition of $3N-3$ eigen modes $k$ with the amplitude $A^k(t)$;
\begin{equation}
\vec{u}_i(t) = \sum_{k=1}^{3N-3} A^k(t) \vec{e}^k_i \quad (i=1,2,...,N). \label{solution2}
\end{equation}
In thermal systems at finite $T>0$, the eigen modes always exchange energy with each other, due to genuine expansion nonlinearities of the potential (and through the thermal bath as well)~\cite{Ashcroft,McGaughey,McGaughey_2004,Turney2_2009}.
We remark that the expansion nonlinearities should be very small in our temperature range, but they still play a role in thermal equilibration of the system.
This fact is reflected in the observation that all the amplitudes, $A^k(t)$ ($k=1,2,...,3N-3$), are a function of time $t$.
As a consequence, the life-times of the modes, which can be measured by the mode energy correlation function~\cite{McGaughey,McGaughey_2004,Turney2_2009}, are finite, although they become very large at low $T$.
This mode mixing is a generic feature for $T>0$ systems, and is active even at very low $T$, as in the present case.
Particularly, in crystals, it is known as phonon-phonon interaction and plays a central role in thermal conduction~\cite{Ashcroft,McGaughey,McGaughey_2004,Turney2_2009}.
In contrast, if Eqs.~(\ref{harmonic}) and~(\ref{solution}) hold, the modes can \textit{not} interact.
Thus the amplitudes $A^k(t)$ of modes are all constant, not depending on $t$, and the lifetimes of modes are infinite.
Note that in Eq.~(\ref{solution}) $A^k(t) = A^k_0 \exp(j \omega^k t)$, and $|A^k(t)| \equiv A^k_0$ is constant.
Therefore, the \textit{strict} definition of harmonic motions, Eqs.~(\ref{harmonic}) and~(\ref{solution}), is never realized for systems at finite $T>0$.

Although strict harmonic vibrations are not realized in thermal systems, the law of equipartition of energy gives a harmonic condition for the amplitude $A^k(t)$ of mode $k$, as follows.
In the harmonic approximation, the potential energy of the system, $\Delta \Phi(t) = \Phi(t)-\Phi^0$, measured from the minimum value $\Phi^0$ is formulated in terms of the dynamical matrix, $({\partial^2 \Phi}/\partial\vec{r}_i \partial \vec{r}_j)^0$~\cite{Ashcroft,McGaughey,McGaughey_2004,Turney2_2009};
\begin{equation}
\begin{aligned}
\Delta \Phi(t) &= \frac{1}{2} \sum_{i=1}^N \sum_{j=1}^N \vec{u}_i(t) \cdot \left(\derrri{\Phi}{\vec{r}_i}{\vec{r}_j}\right)^0 \cdot \vec{u}_j(t),\\
&= \sum_{k=1}^{3N-3} \left[ \frac{{A^k(t)}^2}{2} \sum_{i=1}^N \sum_{j=1}^N \vec{e}^k_i \cdot \left(\derrri{\Phi}{\vec{r}_i}{\vec{r}_j}\right)^0 \cdot \vec{e}^k_j \right],\\
&= \sum_{k=1}^{3N-3} \frac{ \left[{A^k(t)} \omega^k \right]^2}{2} = \sum_{k=1}^{3N-3} E^k_P(t),
\end{aligned} \label{harmonicenergy}
\end{equation}
where $E^k_P(t) = \left[ A^k(t) \omega^k \right]^2/2$ is the potential energy of eigen mode $k$~\cite{McGaughey,McGaughey_2004,Turney2_2009}.
In the harmonic description, the total potential energy $\Delta \Phi$ is therefore the sum of $E^k_P$ over all modes $k$.
When the system is in thermodynamic equilibrium at a temperature $T$, energy equipartition~\cite{Allen1986} implies that a potential energy $T/2$ is distributed equally among eigen modes, so that
\begin{equation}
\begin{aligned}
\left< E^k_P(t) \right> = & \frac{1}{2} T \quad \Longleftrightarrow \quad \left< {A^{k}(t)}^2 \right> = \frac{T}{{(\omega^k)}^2},
\end{aligned} \label{equipartition}
\end{equation}
where, again, $\left< \right>$ denotes the $NVT$ ensemble average (time average over the MD trajectory).
Therefore, the harmonic value of the amplitude is given as
\begin{equation}
\begin{aligned}
\bar{A}_\text{harm}^k := \sqrt{\left< {A^{k}_\text{harm}(t) }^2 \right>} = \frac{\sqrt{T}}{{\omega^k }},
\end{aligned} \label{equipartition2}
\end{equation}
where we explicitly denote the harmonic value by the subscript ``harm".
Thus, we can identify harmonic vibrations for mode $k$, through the relation ${\bar{A}^k}/{\bar{A}_\text{harm}^k}=1$, where
\begin{equation}
\begin{aligned}
\frac{\bar{A}^k}{\bar{A}_\text{harm}^k} := \frac{ \sqrt{\left< {A^{k}(t)}^2 \right>} }{ \sqrt{\left< { A^{k}_\text{harm}(t)}^2 \right>} } = \frac{ \sqrt{\left< { A^{k}(t)}^2 \right>} }{\sqrt{T}/\omega^k },
\end{aligned} \label{requirement}
\end{equation}
and deviations from the harmonic condition indicate the presence of anharmonicities.
It is important to remark that at low temperatures, the expansion nonlinearities are very weak (even if they are still strong enough to induce mode interactions, as we mentioned above), and do not affect the harmonic approximation of the potential energy.
Therefore, if for mode $k$, ${\bar{A}^k}/{\bar{A}_\text{harm}^k} \ne 1$, this implies that the mode is deformed primarily by the cut-off nonlinearities.

Figure~\ref{amplitude} shows the ratio of the calculated amplitude and the harmonic expectation (solid line), $\bar{A}^k/\bar{A}_\text{harm}^k$, versus the eigen frequency $\omega^k$ at the indicated values of $T$, for all investigated cases (glass and crystal).
For potential $1$, many eigen modes $k$ show larger amplitudes than the harmonic expectation, i.e. anharmonic vibrations.
In glass $1$, the low $\omega^k$ modes are anharmonic even at very low $T=10^{-4}$, whereas at higher frequencies, $\omega^k \gtrsim 5$, the modes become harmonic.
On the other hand, in crystal $1$, many modes over the entire range of $\omega^k$ show large anharmonic effects.
Remarkably, those anharmonic vibrations are completely suppressed in potential $2$, where all the modes are consistent with the harmonic expectation value, in both glass $2$ and crystal $2$.
We emphasize that cut-off nonlinearities should exist in both potentials $1$ and $2$, since the second derivatives of the potentials are both non-analytic at $r=r_c$.
What we have demonstrated here is that cut-off nonlinearities are most pronounced in the case of a discontinuity in the interparticle force at $r=r_c$, which indeed makes a visible impact on the thermal vibrations.
As we will see below, the strength of cut-off nonlinearities in each mode $k$ is then determined by the number of pairs of particles that are involved in the mode $k$ and experience the force discontinuity.

{\bf Comparing glass 1 and crystal 1:} Crystal $1$ shows much larger anharmonic amplitudes in many of the normal modes.
As we mentioned in Fig.~\ref{gr0}(b) of the RDF $g(r)$ (Sec.~\ref{secstatic}), all the particles in crystal $1$ have their nearest neighbors located around the cut-off distance $r=r_c=2.5$ (which coincides with the exact lattice position), and in addition, the eigen modes are phonons, i.e. collective, extended modes.
In this situation, many modes have a large number of pairs of particles which experience the discontinuous force causing large cut-off nonlinearities.
On the other hand, in glass $1$, fewer particles have such neighbors located precisely at $r=r_c=2.5$ because of the amorphous structure.
Furthermore, the eigen modes are less extended than regular phonons, exhibiting quasi-localized features, resulting in much smaller cut-off nonlinearities.

{\bf Comparing glasses 1 and 3:} We observe smaller anharmonic amplitudes in glass $3$ than those in glass $1$.
In glass $3$, although even more pairs of particles, located around $r_c=3.0$, are found than those at $r_c=2.5$ in glass $1$, the discontinuity in the interparticle force is smaller by the longer cut-off distance $r_c=3.0$ (see Fig.~\ref{ljpotential}(b)), which leads to the smaller cut-off nonlinearities.

{\bf Comparing crystals 1 and 3:} In crystal $3$, all the modes show values around the harmonic expectation, i.e. the large anharmonic amplitudes observed in crystal $1$ completely disappear.
The discontinuity at $r=r_c=3.0$ in potential $3$ is not located at the lattice points (see $g(r)$ in Fig.~\ref{gr0}(b)), therefore no pairs of particles experience the discontinuity at low $T$s, and no cut-off nonlinearities appear in crystal $3$.

{\bf Low-$\omega^k$ modes in glasses 1 and 3:} One noticeable observation in glasses $1$ and $3$ is that anharmonicities are found only in the low $\omega^k \lesssim 5$ region, with lower $\omega^k$ modes showing larger anharmonic amplitudes, $\bar{A}^k/\bar{A}_\text{harm}^k>1$.
This result is explained by the fact that the lower $\omega^k$ modes involves more particles which experience the force discontinuity to cause cut-off nonlinearities.
As we discussed in Fig.~\ref{length} of $C^k(r_{ij})$ (Sec.~\ref{secharmonic}), the low $\omega^k$ mode is actually quasi-localized~\cite{mazzacurati_1996,Schober_2004,Silbert_2009,Xu_2010}, i.e. although it is localized, it exhibits an extended character with long spatial correlation.
Therefore, the larger cut-off nonlinearities at the lower $\omega^k$ reflect the extended character of the low $\omega^k$ mode.

{\bf Effect of temperature in crystal 1:} Whereas at lower temperatures particle displacements are small enough that interacting pair separations remain close to the initial distance, at higher temperatures thermal vibrations increase in amplitude causing pairs of particles, initially located near the cut-off distance $r=r_c$, to pass through $r_c$ less often - particle separations reside inside or outside the cut-off distance - thereby reducing the effect of the cut-off nonlinearities.
This is clearly observed in crystal $1$, i.e. as $T$ increases, anharmonic amplitudes are strongly suppressed toward the harmonic value, $\bar{A}^k/\bar{A}_\text{harm}^k=1$.
Here we note that expansion nonlinearities are expected to be enhanced by larger vibrational amplitude at higher $T$, however we cannot resolve these effects at the values of $T$ explored here.
We also remark that cut-off nonlinearities have an effect that is opposite to the temperature of the expansion nonlinearities; increasing temperature leads to a suppression of cut-off nonlinearities, while expansion nonlinearities are expected to grow with temperature.

{\bf Effect of temperature in glasses 1 and 3:} The anharmonicities in glasses $1$ and $3$ are quite insensitive to temperature.
Given the fact that systems 1 and 3 share the same type of discontinuity in the force, one might expect that the glassy systems should exhibit similar temperature effects as crystal 1, i.e. suppressed anharmonicities with increasing temperature.
However, due to their amorphous structures, larger thermal vibrations also involve more pairs of particles that experience the force discontinuity, and actually lead to an increase in the anharmonic effects.
These two effects - increasing temperature and a growing number of anharmonic vibrating particle pairs - appear to cancel, and as a result the anharmonicities in glasses $1$ and $3$ are less sensitive to temperature than might be expected.

Finally, in Fig.~\ref{fmsd} we plot the mean-squared-displacement (MSD) of the particles,
\begin{equation}
\begin{aligned}
  \left< \Delta \vec{r}^2(t) \right> = \frac{1}{N} \sum_{i=1}^N \left<
    \left(\vec{r}_i(t)- \vec{r}_i(0) \right)^2 \right>.
\end{aligned} \label{emsd}
\end{equation}
In the harmonic limit, $\left< \Delta \vec{r}^2(t) \right>\propto T$.
To highlight any deviations due to anharmonic behavior, we therefore plot the data normalized by $T$.
We find excellent data collapse for both glass $1$ (Fig.~\ref{fmsd}(a)) and crystal $2$ (Fig.~\ref{fmsd}(c)).
In glass $1$, despite the fact that many low-frequency modes are anharmonic, these anharmonicities are, in a sense, hidden since all the $3N-3$ modes are mixed in $\left< \Delta \vec{r}^2(t) \right>$.
On the other hand, we see clear anharmonic effect in $\left< \Delta \vec{r}^2(t) \right>$ for crystal $1$ at the lowest $T=10^{-3}$ (Fig.~\ref{fmsd}(b)); $\left< \Delta \vec{r}^2(t) \right>$ shows an increase of one order of magnitude that occurs around $t \simeq 3$.
This time-scale coincides with the frequency of the lowest-lying mode, $\omega^k \simeq 2\pi/t \simeq 2$, indicating that the increase in $\left< \Delta \vec{r}^2(t) \right>$ is driven by large anharmonic contributions coming from the low-$\omega^k$ modes.

\begin{figure}[t]
\centering
\includegraphics[width=0.38\textwidth]{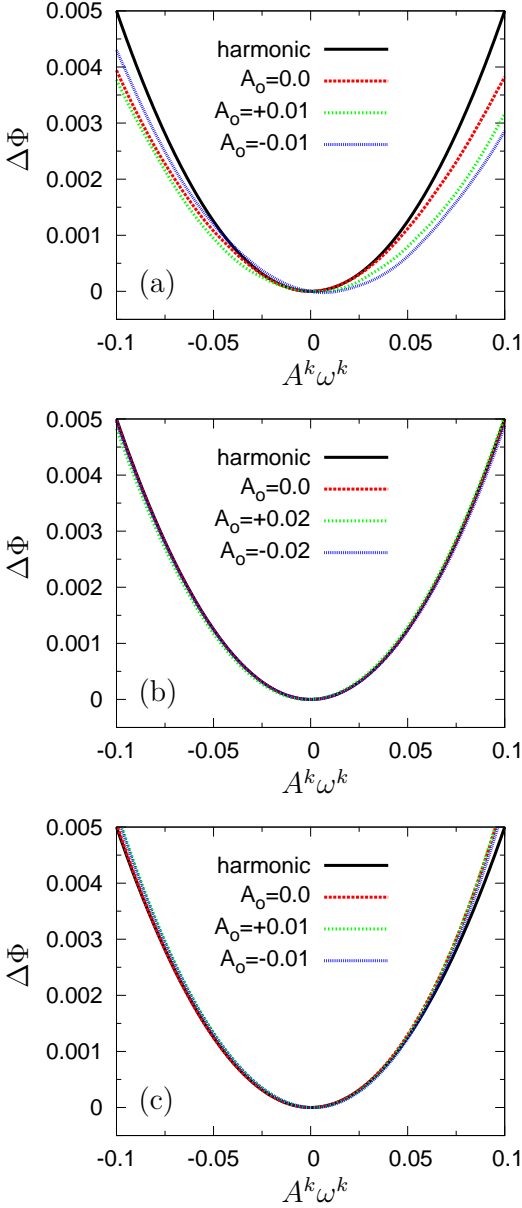}
\vspace*{0mm}
\caption{ 
The potential energy landscape, $\Delta \Phi(A^k,A_\text{o})$, along (a) low $\omega^k=1.13$ and (b) high $\omega^k=8.79$ modes in glass $1$, and (c) low $\omega^k=0.79$ mode in glass $2$.
We \textit{statically} displace the particles from their minimum positions, as $\vec{r}_i = \vec{r}_i^0 + A^k \vec{e}^k_i + A_\text{o} \left( \sum_{l=1,l\neq k}^{3N-3} \vec{e}^l_i/\omega^l \right)$ ($i=1,2,...,N$).
Then we measure the variation of the potential energy from the minimum value, $\Delta \Phi = \Phi-\Phi^0$, as a function of the amplitudes, $A^k,A_\text{o}$.
If the mode $k$ is harmonic, $\Delta \Phi$ shows a parabolic shape, $\Delta \Phi = (A^k \omega^k)^2/2$, independent of $A_\text{o}$, which is indicated by the black solid line.
}
\label{land1}
\end{figure}

\begin{figure}[t]
\centering
\includegraphics[width=0.38\textwidth]{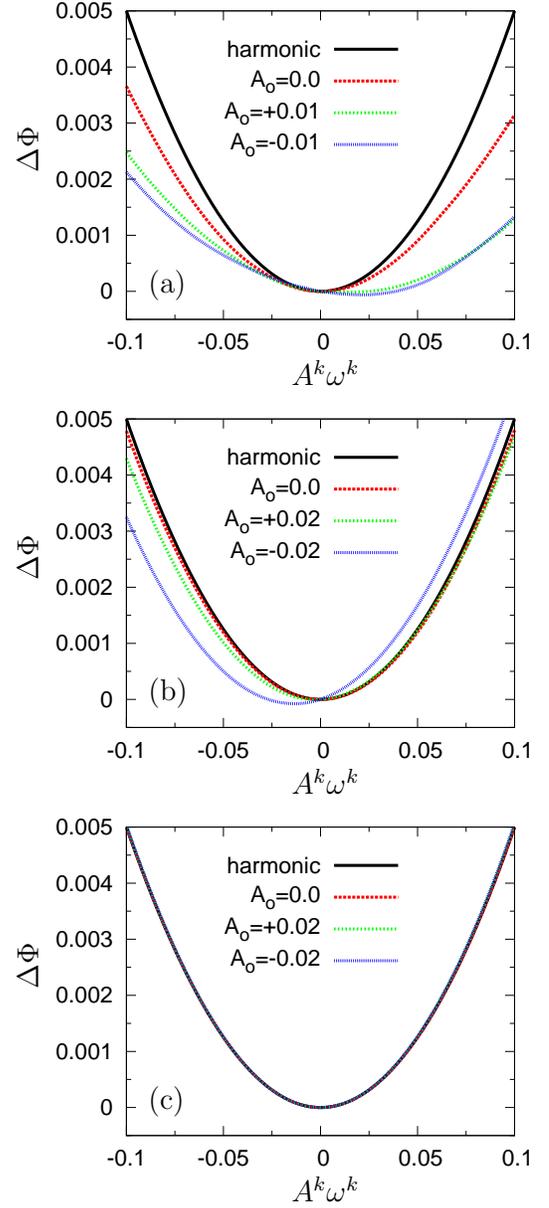}
\vspace*{0mm}
\caption{ 
The potential energy landscape, $\Delta \Phi(A^k,A_\text{o})$, along (a) low $\omega^k=2.19$ and (b) high $\omega^k=8.91$ modes in crystal $1$, and (c) low $\omega^k=2.12$ mode in crystal $2$.
See also the caption of Fig.~\ref{land1}.
}
\label{land2}
\end{figure}

\subsubsection{Potential energy landscape along eigen mode $k$} \label{secpenel} 
In this section, we will interpret the above observation of strongly anharmonic vibrations (Fig.~\ref{amplitude}) in terms of the local geometry of the energy landscape, i.e. variations in the potential energy compared to the parabolic, harmonic limit.
In order to explore the energy landscape close to the minimum, we \textit{statically} displace the particles $\vec{r}_i$ from the energy minimum position $\vec{r}_i^0$ along mode $k$,
\begin{equation}
\vec{r}_i(A^k) = \vec{r}_i^0 + A^k \vec{e}_i^k \quad (i=1,2,...,N).
\end{equation}
$\vec{r}_i(A^k)$ is therefore a function of the amplitude $A^k$ of mode $k$, and we vary $A^k$ continuously from $A^k=0$.
In addition, to probe interactions between modes $k$ and $l \neq k$, we can statically ``excite'' both modes and follow the displacements of particles that we can now express as
\begin{equation}
\vec{r}_i(A^k,A_\text{o}) = \vec{r}_i^0 + A^k \vec{e}^k_i + A_\text{o} \left( \sum_{l=1,l\neq k}^{3N-3} \frac{\vec{e}^l_i}{\omega^l} \right). \label{otherexcitation}
\end{equation}
The value of $A_\text{o}$ determines the extent of the excitations of modes $l \neq k$, and $\vec{r}_i(A^k,A_\text{o})$ is then a function of both $A^k$ and $A_\text{o}$.
Here it is worth to note that if modes $l$ are harmonic in the $T>0$ equilibrium state, their time-averaged amplitudes are $A^l = \sqrt{T}/\omega^l$ as in Eq.~(\ref{equipartition2}), thus the value of $A_\text{o}$ provides a measure of the square root of the temperature, $A_\text{o} = \sqrt{T}$.
The potential energy deviation, $\Delta \Phi = \Phi-\Phi^0$, measured relative to the minimum value $\Phi^0$ (presented in Table~\ref{tsystem}), is then obtained as a function of the two amplitudes, $A^k, A_\text{o}$;
\begin{equation}
\Delta \Phi(A^k,A_\text{o}) = \sum_{\left< i,j \right>} \phi(r_{ij}(A^k,A_\text{o})) - \Phi^0, \label{eqlandscape}
\end{equation}
where $r_{ij}(A^k,A_\text{o}) = |\vec{r}_i(A^k,A_\text{o}) - \vec{r}_j(A^k,A_\text{o})|$, and $\sum_{\left< i,j \right>}$ is the summation over all the pairs of particles $\left< i,j \right>$.

As in Eq.~(\ref{harmonicenergy}), the potential energy landscape along the mode $k$ is formulated in the harmonic approximation as~\cite{Ashcroft,McGaughey,McGaughey_2004,Turney2_2009};
\begin{equation}
\begin{aligned}
\Delta \Phi &= \frac{1}{2} \sum_{i=1}^N \sum_{j=1}^N \vec{u}_i(A^k,A_\text{o}) \cdot \left(\derrri{\Phi}{\vec{r}_i}{\vec{r}_j}\right)^0 \cdot \vec{u}_j(A^k,A_\text{o}),\\
&= \frac{{A^k}^2}{2} \sum_{i=1}^N \sum_{j=1}^N \vec{e}^k_i \cdot \left(\derrri{\Phi}{\vec{r}_i}{\vec{r}_j}\right)^0 \cdot \vec{e}^k_j = \frac{ \left({A^k} \omega^k \right)^2}{2},
\end{aligned} \label{harmonicenergy2}
\end{equation}
where $\vec{u}_i(A^k,A_\text{o}) = \vec{r}_i(A^k,A_\text{o}) - \vec{r}_i^0$.
Thus, if mode $k$ is harmonic, (i) its energy landscape has a parabolic shape as a function of $A^k$, and (ii) it is independent of $A_\text{o}$, i.e. it is independent of the other modes $l \neq k$.
We plot $\Delta \Phi(A^k,A_\text{o})$ for glasses in Fig.~\ref{land1}, and for crystals in Fig.~\ref{land2}, where the harmonic behavior, $\Delta \Phi = (A^k \omega^k)^2/2$, is indicated by the black solid line.

{\bf Energy landscapes in glasses:} Figure~\ref{land1} shows the energy landscapes along, (a) low $\omega^k=1.13$ and (b) high $\omega^k=8.79$ modes in glass $1$.
The energy landscape of the low $\omega^k=1.13$ mode clearly deviates from a parabola shape.
In addition, it is affected by the other vibrational modes $l \neq k$, depending on the value of $A_o$.
We have confirmed that the low $\omega^k=1.13$ mode is affected only by those low-lying modes with, $\omega^k \lesssim 5$ i.e. mode couplings occur only between the lowest $\omega^k \lesssim 5$ modes.
In fact, the high $\omega^k=8.79$ mode maintains a parabolic shape regardless of the other mode excitations with $A_o \neq 0$.
On the other hand, in glass $2$, as seen in Fig.~\ref{land1}(c), even the low $\omega^k=0.79$ mode exhibits a harmonic energy landscape, independent of $A_\text{o}$.
These observations in glasses are consistent with the results of vibrational amplitudes reported in Fig.~\ref{amplitude}(a)-(c).
The discontinuity in the interparticle force in glass $1$ (and also $3$) deforms the energy landscape from a parabolic shape, and induces mode couplings.

{\bf Energy landscapes in crystals:} In Fig.~\ref{land2}, we plot the energy landscapes for crystals.
In crystal $1$, the energy landscape along the low $\omega^k=2.19$ mode is rather flat and also sensitive to the other mode excitations $\l \neq k$, resulting in strongly anharmonic behavior.
The high $\omega^k=8.91$ mode shows a parabola shape for $A_\text{o}=0$, but it is affected by the other mode excitations with $A_\text{o} \neq 0$.
On the other hand, in crystal $2$ (and also $3$), all the modes become harmonic with a parabolic energy landscape.
This result is also consistent with the observation in Fig.~\ref{amplitude}(d)-(f).

One remarkable observation in crystal $1$ is that the mode coupling due to cut-off nonlinearities are found even between low and high $\omega^k$ modes.
This is not the case in glasses $1$ and $3$, where mode couplings occur only between the low $\omega^k$ modes.
The interaction between the modes $k$ and $l$ emerges when those two modes share pairs of particles which experience the force discontinuity at $r=r_c$.
In crystal $1$, if two modes $k,l$ have the same polarization and propagating direction, they share such pairs of particles, regardless of their frequencies, $\omega^k$ and $\omega^l$.
On the contrary, in glasses $1$ and $3$, high $\omega^k$ modes are not extended, consisting of short spatial correlations (see Fig.~\ref{length}), and therefore such modes are less likely to share particles that experience the force discontinuity.
Whereas, it is more likely for the low $\omega^k$ modes, which are extended in character, to have some overlap in their particle vibrations that experience the force discontinuity.

In conclusion to this section, potential $2$ (continuous interparticle force) exhibits no visible anharmonicities in the vibrational amplitudes as well as in the energy landscapes along the modes.
In contrast, there are clear signatures of anharmonic vibrational properties for potentials $1$ and $3$ (discontinuous interparticle forces).
We emphasize again that for potential $2$, although the interparticle force is continuous at $r=r_c$, cut-off nonlinearities are induced by a second derivative which is non-analytic in $r=r_c$.
Hence, our results demonstrate that it is primarily the discontinuity in the interparticle force that enhances the cut-off nonlinearities, deforms the energy landscapes of many modes away from the expected parabola shape, causing leaks in the energy from one mode to other modes, and as a result, induces detectable anharmonic effects.
The number of particles pairs that experience the force discontinuity determines the strength of the cut-off nonlinearities for each mode $k$; more pairs of particles pass through the cut-off point $r=r_c$, stronger anharmonic effects appear.

\subsection{Effects of cut-off nonlinearities on physical quantities} \label{effectofcutoff}
In this Section we study the effects of cut-off nonlinearities on several physical quantities, including the RDF, effective vDOS (Fourier transform of the velocity auto-correlation function), total potential energy, specific heat, and elastic constants.
We will see that anharmonic effects manifest more evidently in quantities which include more detailed information on vibrational states.

\begin{figure}[t]
\centering
\includegraphics[width=0.425\textwidth]{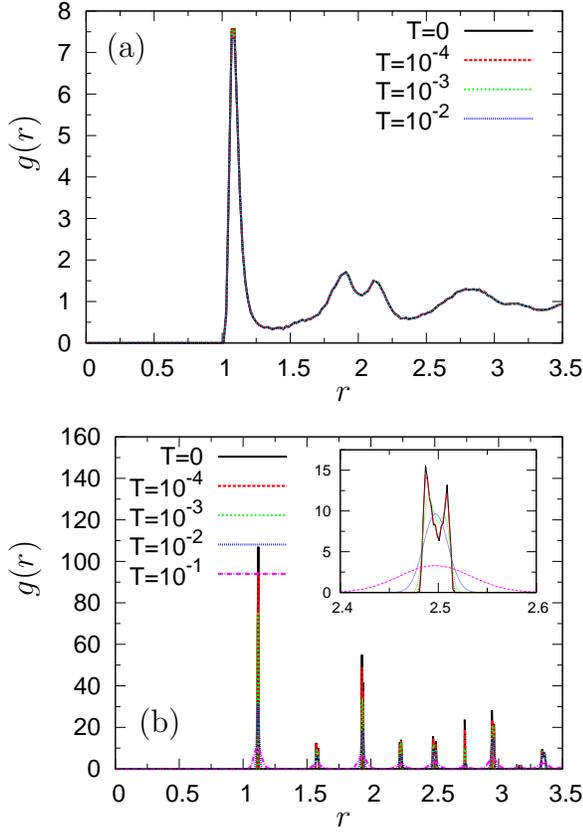}
\vspace*{0mm}
\caption{ 
The RDF $g(r)$ for (a) glass $1$ and (b) crystal $1$ at zero, $T=0$, and finite, $T>0$, temperatures.
The inset to (b) is a close-up around $r=r_c=2.5$.
The data at $T=0$ are same as those presented in Fig.~\ref{gr0}.
}
\label{grt}
\end{figure}

\subsubsection{Radial distribution function (static structure)} \label{secstatict}
In Fig.~\ref{grt}, we compare the $g(r)$ at $T>0$ with that at $T=0$, for potential $1$ (glass $1$ and crystal $1$).
As the temperature increases, particles execute larger thermal vibrations, and the peaks in $g(r)$ decrease in height and broaden.
This temperature effect is clearly observed in the crystal case (Fig.~\ref{grt}(b)).
There is also a decrease of the first peak in glass $1$, although this change is quite small compared to the former case (Fig.~\ref{grt}(a)).
Figure~\ref{grt} demonstrates that as $T$ tends to zero, $g(r)$ smoothly converges to that at $T=0$; $g(r)$ at $T=10^{-4}$ coincides well with that at $T=0$.
In crystal $1$, we observe the single peak around each lattice position at higher $T \ge 10^{-2}$, while this peak splits into two peaks at lower temperature, $T \le 10^{-3}$ (see inset).
These twin peaks are \textit{not} related to the cut-off nonlinearities, but they simply originate from the instability due to the force discontinuity at $r=r_c$, as discussed in Fig.~\ref{gr0}(b) and Sec.~\ref{secstatic}.
Therefore, cut-off nonlinearities (see Figs.~\ref{amplitude} to \ref{land2}) have no effects on $g(r)$.
We draw a similar conclusion for potentials $2$ and $3$.

\begin{figure}[t]
\centering
\includegraphics[width=0.425\textwidth]{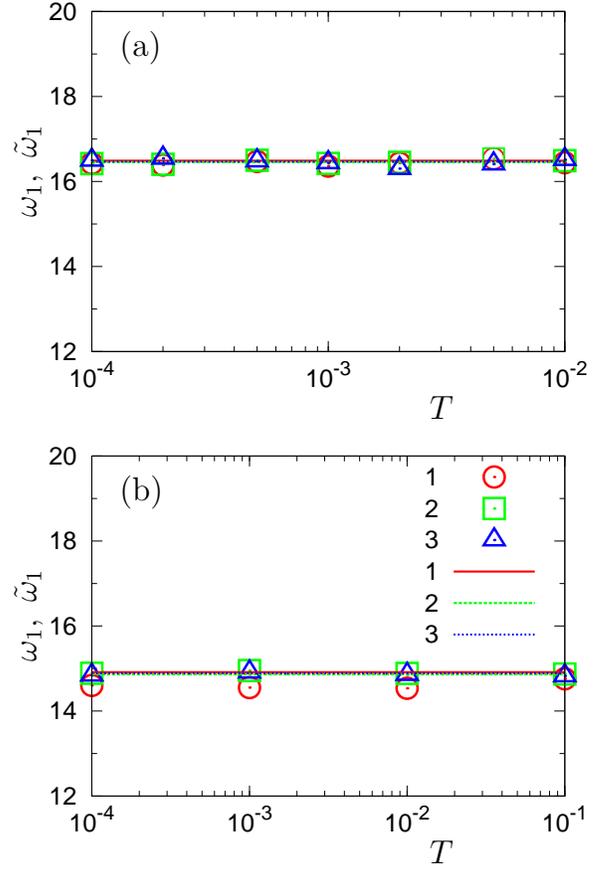}
\vspace*{0mm}
\caption{ 
The first moments, $\omega_1$ and $\tilde{\omega}_1$, of the vDOS, ${d}(\omega)$ and $\tilde{d}(\omega)$, for (a) glasses and (b) crystals.
We plot the temperature $T>0$ dependence of $\tilde{\omega}_1$, by the symbols, and $\omega_1$, i.e. the values in the harmonic limit $T=0$, by the lines.
}
\label{fmom}
\end{figure}

\subsubsection{Vibrational density of states}
Next we look at the effective vDOS, $\tilde{d}(\omega)$, obtained as the Fourier transform, of the velocity auto-correlation function $d(t)$ at $T>0$;
\begin{equation}
\begin{aligned}
d(t) =& \frac{1}{(3N-3)T} \sum_{i=1}^{N} \left< \vec{v}_i(t) \cdot \vec{v}_i(0) \right>, \\
\tilde{d}(\omega) =& \int_0^\infty d(t) \exp(j\omega t) dt,
\end{aligned} \label{vafeq}
\end{equation}
where $\vec{v}_i(t)$ is the velocity of particle $i$. For $T \rightarrow 0$, $\tilde{d}(\omega)$ is expected to converge to the \textit{harmonic} vDOS, $d(\omega)$, obtained from the dynamical matrix, Eq.~(\ref{vdoseq})~\cite{shintani_2008}.
This is tested in Figs.~\ref{dos} and \ref{fmom}.
Figure~\ref{dos} compares $d(\omega)$ (symbols) and $\tilde{d}(\omega)$ at $T=10^{-3}$ (lines), for the three investigated potentials.
In addition, in Fig.~\ref{fmom}, we compare the first moments, $\omega_1$ (lines) and $\tilde{\omega}_1$ (symbols), for $d(\omega)$ and $\tilde{d}(\omega)$, respectively~\cite{Ikeda_2013};
\begin{equation}
\begin{aligned}
\omega_1 &= \int_0^{\infty} \omega d(\omega) d\omega,\qquad \tilde{\omega}_1 = \int_0^{\infty} \omega \tilde{d}(\omega) d\omega.
\end{aligned} \label{vafeq2}
\end{equation}

In Fig.~\ref{fmom}(b), we see small differences between $d(\omega)$ and $\tilde{d}(\omega)$ for crystal $1$; the value $\tilde{\omega}_1$ at finite $T>0$ is slightly lower than $\omega_1$ at $T=0$.
This is caused by the anharmonicities observed in Figs.~\ref{amplitude}(d) and~\ref{land2}(a),(b).
Although crystal $1$ shows large cut-off nonlinearities for many eigen modes, they have only a small effect on the vDOS, shifting the mode frequencies to slightly lower values.
Except for these small differences in crystal $1$, Figs.~\ref{dos} and \ref{fmom} demonstrate good agreement between $d(\omega)$ and $\tilde{d}(\omega)$ over the temperature regime studied.
Therefore, we conclude that the vDOS and the actual values of the mode frequencies $\omega^k$ are insensitive to the cut-off nonlinearities.

\begin{figure}[t]
\centering
\includegraphics[width=0.425\textwidth]{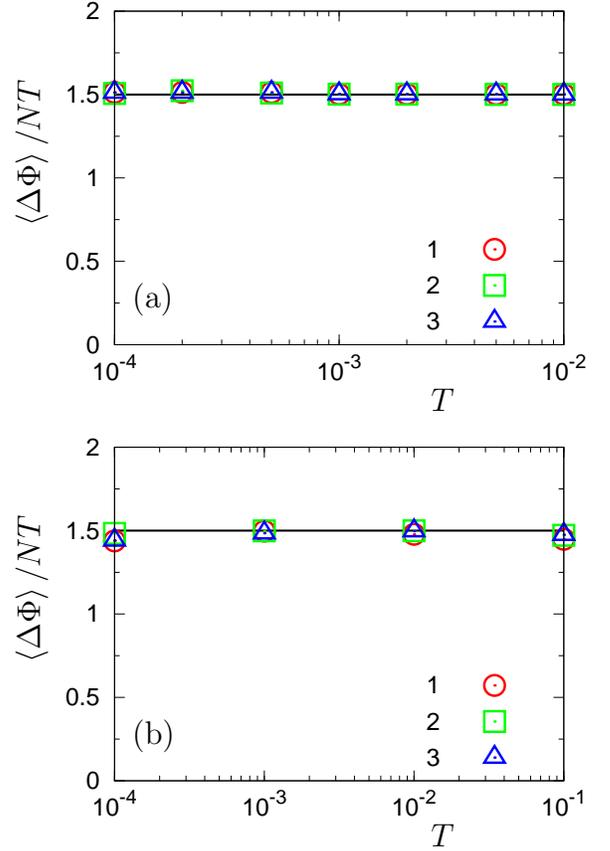}
\vspace*{0mm}
\caption{
The temperature, $T$, dependence of the scaled potential energy, $\left< \Delta \Phi \right>/NT$, for (a) glasses and (b) crystals.
$\left< \Delta \Phi \right>$ is the total potential energy measured from the minimum value $\Phi^0$ (see Eq.~(\ref{cveq})).
If thermal vibrations are harmonic, $\left< \Delta \Phi \right>/NT = 3/2$ - shown as the horizontal solid line.
} 
\label{potential}
\end{figure}

\begin{figure}[t]
\centering
\includegraphics[width=0.425\textwidth]{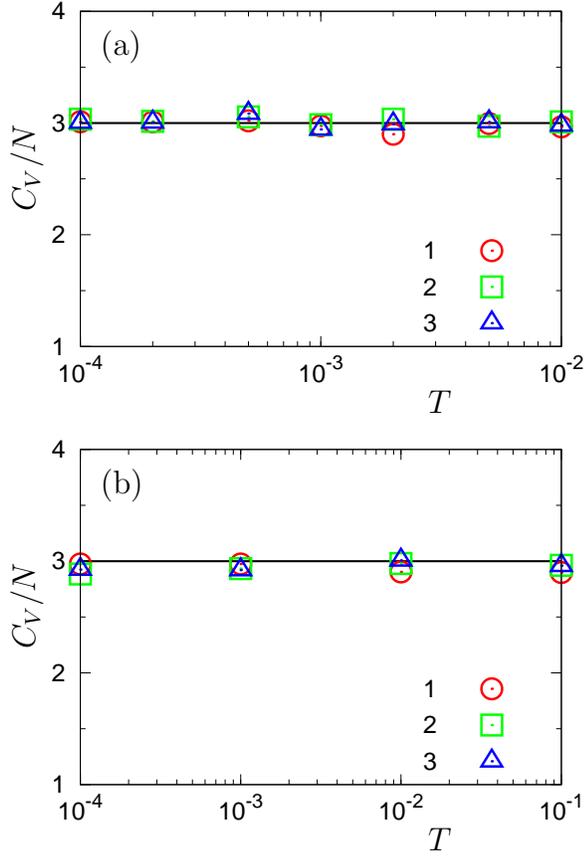}
\vspace*{0mm}
\caption{
The temperature, $T$, dependence of the specific heat per particle, $C_V/N$ (see Eq.~(\ref{cveq})), for (a) glasses and (b) crystals.
If thermal vibrations are harmonic, $C_V/N=3$ - shown as the horizontal solid line.
}
\label{cvr}
\end{figure}

\begin{figure}[t]
\centering
\includegraphics[width=0.425\textwidth]{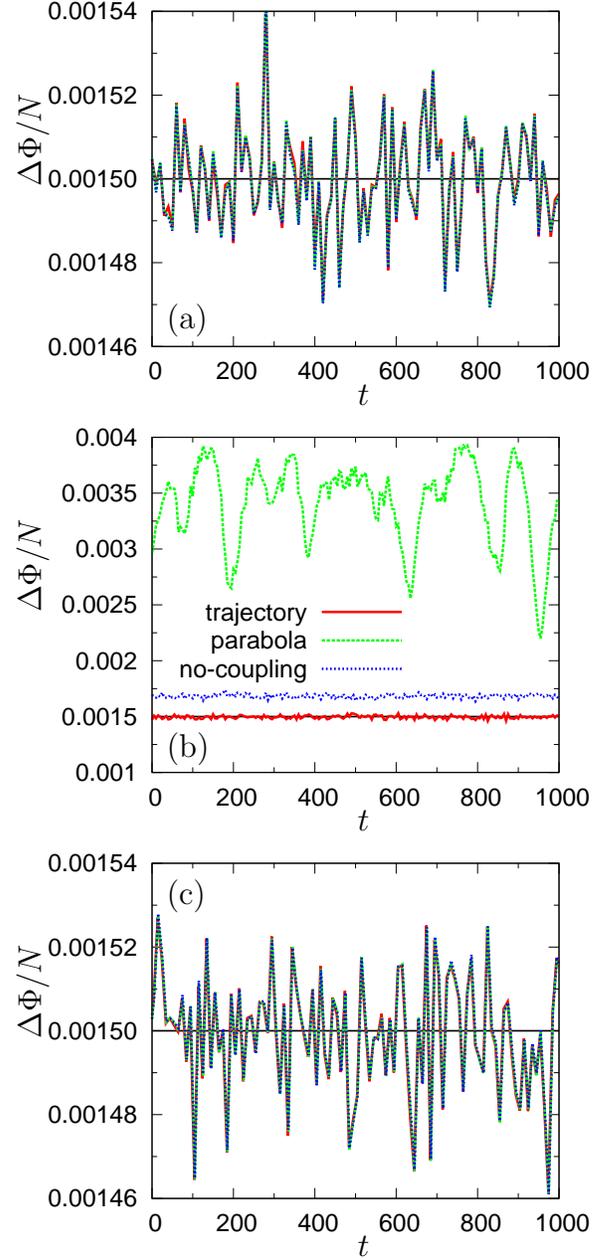}
\vspace*{0mm}
\caption{
Time dependence of the potential energy per particle, $\Delta \Phi(t)/N$, for (a) glass $1$, (b) crystal $1$, and (c) crystal $2$, at $T=10^{-3}$.
Comparison between the true value from the MD trajectory (red solid line), parabola value $\Delta \Phi_\text{parabola}$ (green dashed line), and no-coupling value $\Delta \Phi_\text{no-coupling}$ (blue dotted line) (see Eq.~(\ref{potharmeq})).
Data are plotted every time interval $\Delta t = 10$.
The horizontal black solid line indicates the harmonic value, $\Delta \Phi/N = (3/2)T = 1.5 \times 10^{-3}$.
}
\label{timepotential}
\end{figure}

\subsubsection{Potential energy and specific heat}
We next study the total potential energy deviation $\left< \Delta \Phi \right>$ (measured from the minimum value $\Phi^0$) and the specific heat $C_V$, which are calculated as an ensemble average over the ($NVT$) MD trajectories as
\begin{equation}
\begin{aligned}
\left< \Delta \Phi \right> =& \left< \sum_{<i,j>} \phi(r_{ij}) \right> - \Phi^0, \\
C_V =& \frac{\left< (E -\left< E \right> )^2 \right>}{T^2} = \frac{\left< \delta E^2 \right>}{T^2},
\end{aligned} \label{cveq}
\end{equation}
where $E=K+\Phi$ ($K=\sum_{i=1}^N \vec{v}_i^2/2$ is the kinetic energy) is the total energy of the system, and the $\delta E = E -\left< E \right>$ are the associated fluctuations.

In Figs.~\ref{potential} and~\ref{cvr} we show the $T$-dependences of $\left< \Delta \Phi \right>$ and $C_V$, respectively.
Equipartition of energy directly provides the harmonic values for $\left< \Delta \Phi \right> = (3N-3)T/2 \simeq (3/2)NT$ and $C_V=3N-3\simeq 3N$, which are indicated by horizontal solid lines in the figures.
We clearly see that $\left< \Delta \Phi \right>$ and $C_V$ coincide with these values in the investigated $T$-range, for all the glasses and crystals, which is not surprising.
Indeed, the potential energy (and the specific heat) is related to the eigen frequencies $\omega^k$ in the harmonic limit (Eq.~(\ref{harmonicenergy})), and the values of $\omega^k$ and $g(\omega)$ are not affected by cut-off anharmonicities as shown in Figs.~\ref{dos} and \ref{fmom}.

It is, however, still surprising that crystal $1$ retains strongly harmonic character in the values of $\left< \Delta \Phi \right>$ and $C_V$, despite strong anharmonic effects which largely deform the energy landscapes of the vibrational modes as shown in Figs.~\ref{land2}(a) and (b).
To further elucidate this point, we calculated the potential energy by using ``parabolic'' and ``no-coupling'' formulations;
\begin{equation}
\begin{aligned}
\Delta \Phi_\text{parabolic} (t) =& \sum_{k=1}^{3N-3} \frac{ \left[ \omega^k A^k(t) \right]^2 }{2}, \\
\Delta \Phi_\text{no-coupling} (t) =& \sum_{k=1}^{3N-3} \Delta \Phi (A^k(t),A_\text{o}\equiv 0), \label{potharmeq}
\end{aligned}
\end{equation}
where the value of $A^k(t)$ is the amplitude of mode $k$ defined in Eq.~(\ref{ampleq}), and $\Delta \Phi (A^k,A_\text{o})$ is the energy landscape defined in Eq.~(\ref{eqlandscape}).
We recall that, as shown in Figs.~\ref{land1} and~\ref{land2} (Sec.~\ref{secpenel}), cut-off nonlinearities (i) deform the energy landscape from a parabolic shape and (ii) induce couplings between the modes.
$\Delta \Phi_\text{parabolic}$ disregards both effects, while $\Delta \Phi_\text{no-coupling}$ includes (i) but not (ii).

In Fig.~\ref{timepotential}, we compare the time dependencies of those two quantities ($\Delta \Phi_\text{parabola}(t)$ and $\Delta \Phi_\text{no-coupling}(t)$) with that of the true value evaluated directly from the MD trajectory $\Delta \Phi(t)$.
We see that $\Delta \Phi_\text{parabola}(t)$ and $\Delta \Phi_\text{no-coupling}(t)$ coincide well with the true $\Delta \Phi(t)$ in glass $1$ (Fig.~\ref{timepotential}(a)).
In glass $1$, even $\Delta \Phi_\text{parabola}(t)$ works well to catch the instantaneous value of $\Delta \Phi(t)$, because anharmonic vibrations appear only in the low $\omega^k \lesssim 5$ regime, and their effects are averaged out by the summation over all the modes ($k=1$ to $3N-3$).

On the other hand, in crystal $1$, the parabolic formulation $\Delta \Phi_\text{parabola}(t)$ completely misses the true value $\Delta \Phi(t)$ (Fig.~\ref{timepotential}(b)).
This is obviously due to the large anharmonic amplitudes, $A^k(t)$.
The no-coupling formulation $\Delta \Phi_\text{no-coupling}(t)$, in contrast, behaves much better, since it takes into account the flattening of the energy landscapes due to large vibrational amplitudes, but it still deviates from the true value $\Delta \Phi(t)$.
Thus, both the flattening of the energy landscapes and the mode-couplings contribute to the total potential energy $\Delta \Phi(t)$.
As a final observation, as demonstrated by the MD trajectory value in Fig.~\ref{timepotential}(b) (red solid line), $\Delta \Phi(t)$ fluctuates around the harmonic value $(3/2)NT$.
Although the dynamics projected on the eigen modes defined at zero temperature does not follow the equipartition of energy, these low temperatures are enough to actually smooth out the energy landscape and result in an effectively harmonic behavior for the energy, in the spirit of the self-consistent phonon theory~\cite{Klein_1972}.
Finally, as can be expected, crystal $2$ shows good agreement between the three time histories of $\Delta \Phi(t), \Delta \Phi_\text{parabola}(t), \Delta \Phi_\text{no-coupling}(t)$ (Fig.~\ref{timepotential}(c)).

\begin{figure*}[t]
\centering
\includegraphics[width=0.9\textwidth]{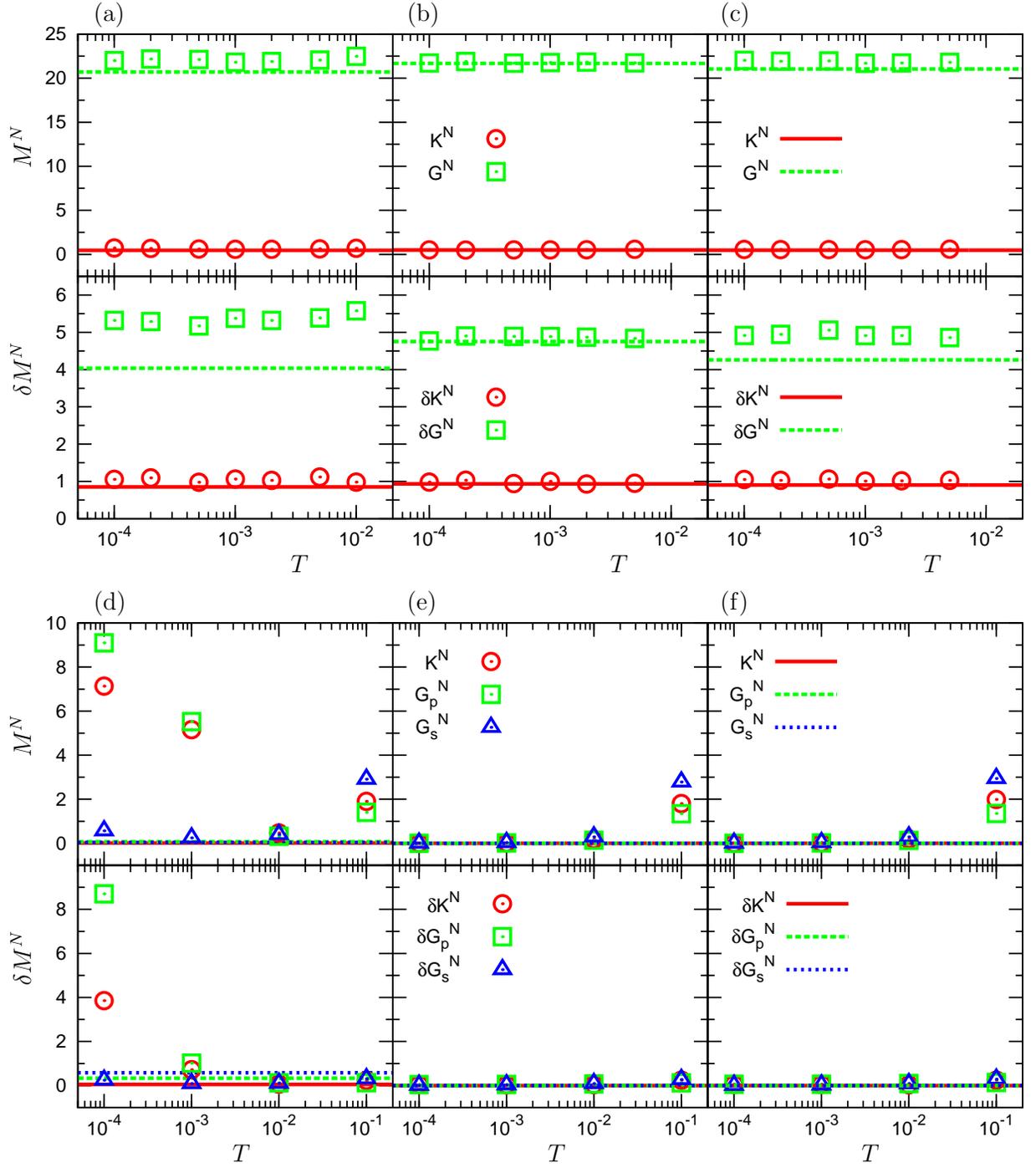}
\vspace*{0mm}
\caption{
Dependence of non-affine moduli $M^{Nm}$ (Eq.~(\ref{moduluseq2})) on temperature; (a) glass~$1$, (b) glass~$2$, (c) glass~$3$, (d) crystal~$1$, (e) crystal~$2$, and (f) crystal~$3$.
We plot the average (global) value $M^N$ and the standard deviation $\delta M^N$ of the distribution function $P(M^{Nm})$ (see Eq.~(\ref{sds})).
Values for the bulk $K^N$ (red circle), pure shear $G_p^N$ (green square), and simple shear $G_s^N$ (blue triangle) moduli are presented.
In glasses (a)-(c), the values for the two shear moduli, $G_p^N$ and $G_s^N$, coincide with each other, which is denoted by $G^N$ (green square).
The lines indicate values in the harmonic approximation, $T = 0$ (Eq.~(\ref{moduluseq3})).
Details of the calculations of elastic moduli are given in Appendix~\ref{methodmodulus}.
} \label{nonaffine}
\end{figure*}

\subsubsection{Global and local elastic moduli} \label{resultmodulus}
As we have demonstrated above, effects of the cut-off nonlinearities are not particularly noticeable on the RDF, vDOS, total potential energy, and specific heat.
In contrast, we show here that the elastic constants are relatively sensitive to cut-off nonlinearities.
We computed the elastic moduli at finite $T>0$, and at $T=0$ in the harmonic approximation, by using the methods described in Appendix~\ref{methodmodulus}.
In the present study, we focus on the non-affine components, $M^{Nm}=K^{Nm}$, $G_p^{Nm}$, and $G_s^{Nm}$ of the bulk, pure shear, and simple shear moduli, respectively, since anharmonic vibrations are directly reflected in these non-affine components.
Here $m$ indicates coarse-grained cubic domains of size much smaller than the simulation box size, as detailed in Appendix~\ref{methodmodulus}.
The affine components, $M^{Am}=K^{Am}, G_p^{Am}, G_s^{Am}$, are determined rather by static structural properties, which are insensitive to anharmonicities as seen in $g(r)$ of Fig.~\ref{grt}.
We have indeed confirmed that the affine components are not influenced by cut-off nonlinearities.

In Figure~\ref{nonaffine} we present the average (global) value $M^N$ and the fluctuations (standard deviation) $\delta M^N$ of the probability distributions $P(M^{Nm})$ (see Eq.~(\ref{sds})), as functions of $T$, for each considered glass and crystal.
$\delta M^N$ is a measure of the elastic heterogeneity.
In the glasses (Fig.~\ref{nonaffine}(a)-(c)), both values of $M^N$ and $\delta M^N$ are insensitive to $T$ variations over the temperature regime studied.
Glass $1$ clearly shows a disagreement between the finite $T>0$ value and the harmonic value ($T=0$), whereas good agreement is observed in glass $2$.
The disagreement in glass $1$ is caused by the cut-off nonlinearities, which are enhanced by the discontinuous force at $r=r_c$.
We note that the difference between the $T>0$ and $T=0$ values shows up more clearly in the shear modulus $G^N, \delta G^N$ than the bulk modulus $K^N, \delta K^N$, since the shear modulus expresses a larger non-affine component~\cite{Wittmer_2002,Tanguy_2002}.
In addition, $\delta M^N$ shows larger discrepancies than the global $M^N$, indicating that local quantities are more sensitive to anharmonicities than global values.
The cut-off nonlinearities cause the $T>0$ value of $\delta M^N$ to increase away from the $T=0$ value, i.e. the moduli distributions become more heterogeneous.
Indeed, in disordered amorphous structures, particles feel the force discontinuity at $r=r_c$ randomly in space.
Cut-off nonlinearities therefore induce spatially heterogeneous effects.

Glass $3$ also shows disagreement between the $T=0$ and $T>0$ values of $M^N, \delta M^N$, albeit a smaller difference than that for glass $1$.
In glass $3$, cut-off nonlinearities are smaller, and have a smaller effect on the elastic moduli.
In both glasses $1$ and $3$, the disagreement persists down to the lowest temperature, $T =10^{-4}$, which is three orders of magnitude lower than $T_g \simeq 0.4$.
This observation therefore poses a caveat on the value $T=10^{-4}$, which is sometimes considered as a temperature where the harmonic limit is recovered.
As we have shown above, this peculiar behavior is due to the fact that cut-off nonlinearities persist to the very-low-$T$ regime, contrary to usual expansion nonlinearities which vanish at small temperatures.

Turning to the crystals (Fig.~\ref{nonaffine}(d)-(f)), we realize that the cut-off nonlinearities produce completely unexpected and anomalously large values of the elastic moduli in crystal $1$.
At low-$T$, the mechanical response of crystals is described only in terms of the affine modulus~\cite{elastictheory}.
Additionally, the ordered lattice structure of a uniform crystal should lead to a spatially distribution in the elastic moduli that is homogeneous.
As a result, we expect that not only should the (total) non-affine modulus vanish, $M^N=0$, but so should the corresponding fluctuations, $\delta M^N=0$, i.e. spatially, all the local values of non-affine moduli, $M^{Nm} \equiv 0$.
In fact, this is indeed true for crystals $2$ and $3$, as demonstrated in Fig.~\ref{nonaffine}(e) and (f).
Note that the increase of $M^N$ at $T=10^{-1}$ is caused by expansion nonlinearities due to large thermal vibrations.
However, crystal $1$ clearly shows large deviations in $M^N>0$ and $\delta M^N>0$ compared to the zero expected value.
For the case of the shear moduli fluctuations, we note that although the values are small, $\delta G_p^N$ and $\delta G_s^N$ assume finite values even at $T=0$.
This comes from the fact that particles are slightly displaced from the exact lattice positions due to the force discontinuity at $r=r_c=2.5$, i.e. crystal $1$ is slightly inhomogeneous, as demonstrated by the $g(r)$ of Fig.~\ref{gr0}(b).

We underline here an interesting observation in the case of crystal $1$; the bulk $K^N$ and pure shear $G_p^N$ moduli are strongly affected by the cut-off nonlinearities, whereas the simple shear modulus $G_s^N$ is much less sensitive.
This result indicates that the simple shear phonon modes, i.e. transverse phonons in the direction $[1,0,0]$, tend to avoid the force discontinuity at $r=r_c$ and are less affected by the cut-off nonlinearities, compared to phonons propagating in other directions.

To sum up this section, anharmonicities due to cut-off nonlinearities cause relatively large effects on the elastic moduli.
In the glassy states, systematic deviations up to $10$-$20\%$ persist in the moduli and their fluctuations over the entire temperature range.
While, for the crystals, anharmonic effects can cause significant deviations even at the lowest temperatures.
None of these anharmonic effects are evident in the RDF, vDOS, potential energy, and specific heat.
These facts point to the conclusion that the strength of the anharmonic effects on a physical quantity seems to be determined by the total amount of detailed information regarding vibrational excitations it includes.
Indeed, the vDOS, potential energy, specific heat, only involve functions of the eigen frequencies $\omega^k$ and are only mildly sensitive to the cut-off nonlinearities.
In contrast, as it is evident in the harmonic equation, Eq.~(\ref{moduluseq3}), the non-affine moduli are described in terms of both the eigen frequencies $\omega^k$ and eigen vectors $\vec{e}^k$, i.e. they include a complete information on the structure of the vibrational modes.
Also note that eigen vectors are more strongly affected by the nonlinearities than the eigen frequencies~\cite{Goodrich2_2014}.

\section{Summary} \label{summary}
In the present paper, we have studied the effects of cut-off nonlinearities induced by the interaction potential cut-off on low-temperature thermal vibrations in glasses and crystals.
It is common practice in computer simulations to truncate the (long-range) interaction potential at some cut-off distance $r=r_c$.  Here, we have focused on three specific parameterizations of the traditional LJ potential, as listed in Table~\ref{tsystem} and Fig.~\ref{ljpotential}; (i) potential $1$: shifted potential $\phi_{SP}(r)$ with $r_c=2.5$, discontinuous 1st and 2nd derivatives, (ii) potential $2$: shifted-force potential $\phi_{SF}(r)$ with $r_c=2.5$, discontinuous 2nd derivative, and (iii) potential $3$: shifted potential $\phi_{SP}(r)$ with $r_c=3.0$, discontinuous 1st and 2nd derivatives.
While these three truncated potentials do not show any noticeable differences in the static structure (Fig.~\ref{gr0}) and vibrational states (Figs.~\ref{dos} and ~\ref{pr}) in the (harmonic) limit $T=0$, differences become apparent in thermal vibrations at finite temperatures $T>0$ (Figs.~\ref{amplitude} to \ref{land2}), which originate from the cut-off nonlinearities.
The truncation causes the potential to be non-analytic at the cut-off, and the harmonic equations of motions, Eq.~(\ref{harmonic}), do \textit{not} account for this singularity.
Thus, when some pairs of particles pass through $r_c$, the harmonic description breaks down, leading to anharmonicities in physical observables.
The cut-off nonlinearities are distinct from the usual expansion nonlinearities, which become more prominent at higher temperatures, that come from neglecting higher order terms in the Taylor expansion of the potential.

We have demonstrated that it is primarily the discontinuity in the interparticle force (the first derivative of the potential) at $r=r_c$ that enhances the cut-off nonlinearities.
The force discontinuity deforms the potential energy landscapes of many eigen modes away from the harmonic parabolic shape, as shown in Figs.~\ref{land1} and~\ref{land2}.
In addition, when normal modes share the same pairs of particles experiencing the force discontinuity, they exchange or leak energy, leading to further deformation of their energy landscapes due to the modes coupling.
These effects can be quantified in terms of the vibrational amplitudes shown in Fig.~\ref{amplitude}.
For crystal $1$, where the force discontinuity is located at $r_c=2.5$, which also corresponds to the nearest-neighbor spacing, the implication of this are particularly magnified precisely because the distribution of particle separations is a delta function at $r_c$ itself.
As a result, most, if not all, particle pairs participating in normal mode vibrations experience the force discontinuity at the same time, leading to the observed enhanced anharmonic effects.
On the contrary, even though cut-off nonlinearities also exist in glass $2$ and crystal $2$, where the force is continuous at $r_c$, anharmonic effects were largely suppressed and barely detectable in the vibrational amplitudes and energy landscape features.

We also studied the effects of cut-off nonlinearities on several physical quantities, including the RDF, vDOS, potential energy, specific heat, and elastic moduli (Figs.~\ref{grt} to~\ref{nonaffine}).
We showed that the RDF, vDOS, potential energy, and specific heat are resilient to cut-off effects, whereas the elastic moduli are rather sensitive to the anharmonic nature of the normal modes.
Indeed, the non-affine components of the elastic moduli, computed within the harmonic approximation ($T=0$), are determined by both the eigen frequencies $\omega^k$ and crucially the eigen vectors $\vec{e}^k$ (see Eq.~(\ref{moduluseq3})), i.e. they include complete information on the structure of the vibrational excitations.
Therefore, it is understandable that anharmonic effects have a relatively large impact on the elastic moduli.
Particularly, the large anharmonic vibrations in crystal $1$ results in unphysical trends in the elastic moduli.
In contrast, the vDOS, potential energy, and specific heat, which include only the eigen frequency information in the harmonic limit, are insensitive to anharmonicities.

In contrast to expansion nonlinearities, cut-off nonlinearities persist in the very-low-temperature regime, $T=10^{-4} \ll T_g,\ T_m$.
This is especially the case for crystal $1$, where we have found that cut-off nonlinearities actually become more prominent with decreasing temperature.
As a result, the elastic moduli of glasses $1,3$ and crystal $1$ do \textit{not} coincide with the harmonic values even at $T=10^{-4}$, where we normally expect the harmonic description to be valid.
Even though thermal displacements are very small, $\Delta r \sim \sqrt{T}$, pairs of particles with $r=r_c$ feel the force discontinuity, causing the anharmonicities.
This counter-intuitive behavior of the cut-off nonlinearities is obviously at odds with our understanding of expansion nonlinearities, which vanish in the low temperature limit.

\begin{figure}[t]
\centering
\includegraphics[width=0.425\textwidth]{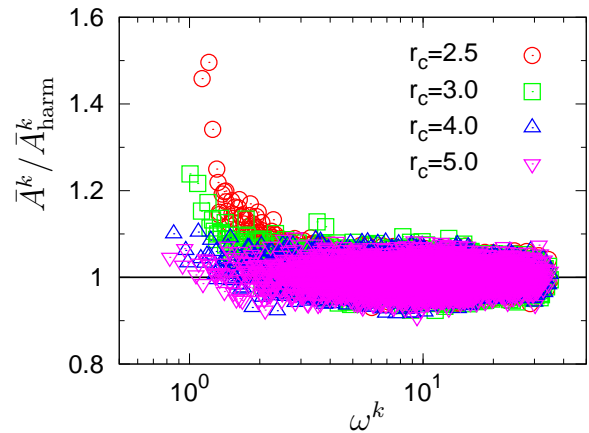}
\vspace*{0mm}
\caption{ 
Amplitude $\bar{A}^k$ versus eigen frequency $\omega^k$ for the shifted potential $\phi_{SP}(r)$ with different cut-off distances, $r_c=2.5,\ 3.0,\ 4.0$, and $5.0$.
The configuration is glass, and the temperature is $T=10^{-3}$.
Note that $r_c=2.5$ and $3.0$ correspond to glasses $1$ and $3$, respectively.
See also the caption of Fig.~\ref{amplitude}.
} \label{depcutoff}
\end{figure}

\section{Conclusions} \label{conclusion} 
We conclude with a few observations.
The factor ultimately determining the strength of the cut-off non-linearities is the number of interacting particles that experience the cut-off discontinuity or, more specifically, the number of inter-particles distances close to the cut-off distance.
This in turn depends on the physics of the system under study.
For the case of the (traditional) 12-6 LJ potential explored here, we found strong discrepancies for crystal $1$ because of the commensurability between the cut-off distance and the choice of the lattice environment (that was tuned according to the density).
The anharmonicities arise because particles fall out of contact with one another during thermal vibrations.
Thus, from an interaction point of view the potential appears to behave one-sided for particle pairs separated by the cut-off distance.
This situation is similar to models of granular materials and jammed particulate packings, where particles interact through one-sided, finite-range (contact/cut-off), potentials~\cite{Schreck_2011,Bertrand_2014,Schreck_2014,Ikeda_2013,Goodrich_2014,Goodrich2_2014}.

Also, for studies at low temperature, in or close to the harmonic regime, one has to be extremely careful with the implementation of the potential truncation.
A cut-off distance $r_c=2.5\sigma$ is often employed in computer simulations involving LJ-type potentials~\cite{Allen1986,Xu_2012,Toxvaerd_2011}.
However, as we have seen in the present work, this value may not be suitable to study low-temperature vibrational properties for particular systems, as it can result in inconsistencies between the harmonic value $T=0$ and the low temperature values.
Though we remark here that in the case of the soft-sphere potential, $\phi_{SC}(r) = \epsilon ( {\sigma}/{r} )^{12}$, which is popular repulsive potential employed in many numerical works~\cite{Allen1986,bernu_1985}, the cut-off discontinuity is significantly smaller than for the 12-6 LJ potential.
In this case, the value of $\phi_{SC}(r)$ at $r_c=2.5$ is $1.7\times 10^{-3}\%$ of $\epsilon$, which is much smaller than that of the LJ potential, $1.6\%$ of $\epsilon$.
In our previous works \cite{Mizuno2_2013,Mizuno_2014}, we employed $r_c=2.5$ in $\phi_{SC}(r)$, which was large enough to study both elastic constants and sound and thermal transport, without any visible cut-off nonlinearities.
Here, we have observed that physical quantities that depend only on the structural arrangement of the particles and/or just the eigen frequencies, such as the RDF, vDOS, potential energy, specific heat, are not appreciably sensitive to the cut-off effects.
However, cut-off nonlinearities may be apparent in others quantities that depend explicitly on the eigen vectors, or particle polarization vectors of the individual particles, such as mechanical properties, for example the elastic moduli.
Thus, the apparent validity of the harmonic regime can depend on which observable is considered~\cite{Goodrich2_2014,Goodrich_2014}.
Low frequency modes are more influenced by the cut-off nonlinearities, so that thermal transport properties are also likely to be affected in a significant manner.

Ideally, we would like to simulate a LJ system without any cut-off of the potential, or one with a long cut-off distance.
Indeed, the cut-off nonlinearities can be suppressed by employing longer cut-offs, thereby reducing the discontinuity at $r=r_c$.
In Fig.~\ref{depcutoff} we plot the vibrational amplitudes $\bar{A}^k$ at the indicated values of $r_c=2.5,\ 3.0,\ 4.0$, and $5.0$, for the glass configuration at temperature $T=10^{-3}$.
From these data, we see that the values of $r_c=4.0$ or $5.0$ can be sufficient to suppress the anharmonic effects in the present LJ system.
However, we have shown that it is the discontinuity in the first derivative of the potential that has a dominant influence.
In addition, we have also confirmed that the vibrational modes are practically identical in the case of long cut-offs ($r_c = 4.0, 5.0$) and shifted-force potential, i.e. force-shifting procedure practically does not alter the vibrational modes.
Therefore, it is recommended that the interparticle force (the first-derivative of the potential) be made continuous at $r=r_c$, as proposed in Refs.~\cite{Toxvaerd_2011,Voigtmann_2009}, to avoid unwanted anharmonic effects without increasing the computational costs by using a large cut off distance.

On the other side, for those potentials that are intrinsically truncated (e.g. soft elastic particles), systems are likely to exhibit nonlinearities~\cite{Schreck_2011,Bertrand_2014,Schreck_2014,Ikeda_2013,Goodrich_2014,Goodrich2_2014} that can strongly influence mechanical properties in athermal or low-temperature states.
Thence, it is particularly important for such systems, that efforts, which extrapolate transport and mechanical properties into the thermal regime purely from structural data or the static harmonic formulation, should be treated with caution.
For instance, predicting material properties such as modes of failure could potentially lead to unexpected catastrophic effects when such results be employed in the design of critical devices.

\begin{acknowledgments}
We thank H.~R.~Schober, M.~Sperl, Th.~Voigtmann, A.~Ikeda, and C.~S.~O'Hern for helpful discussions and useful comments.
\end{acknowledgments}

\appendix

\begin{table}[t]
\centering
\renewcommand{\arraystretch}{1.1}
\begin{tabular}{c|c|ccc|ccc|ccc}
\hline
\hline
           & $T$ & $K$ & $K^A$ & $K^N$ & $G_p$ & $G_p^A$ & $G_p^N$ & $G_s$ & $G_s^A$ & $G_s^N$ \\
\hline
Glass$2$   & $0$       & $60.7$ & $61.2$ & $0.5$ & $14.0$ & $35.7$ & $21.7$ & $14.0$ & $35.7$ & $21.7$ \\
\cline{2-11}
           & $10^{-3}$ & $60.7$ & $61.2$ & $0.5$ & $13.9$ & $35.7$ & $21.8$ & $13.9$ & $35.7$ & $21.8$ \\
\hline
\hline
Crystal$2$ & $0$       & $43.4$ & $43.4$ & $0.0$ & $14.4$ & $14.4$ & $0.0$ & $38.1$ & $38.1$ & $0.0$ \\
\cline{2-11}
           & $10^{-3}$ & $43.7$ & $43.7$ & $0.0$ & $14.7$ & $14.7$ & $0.0$ & $38.2$ & $38.2$ & $0.0$ \\
\hline
\hline
\end{tabular}
\caption{
The average (global) elastic moduli, $K$, $G_p$, and $G_s$, of glass $2$ and crystal $2$.
The temperature is $T=0$ and $10^{-3}$.
We also present values of the affine and non-affine moduli.
Note that for glass $2$, $G_p=G_s$.
} \label{tmoduli}
\end{table}

\section{Measuring finite temperature $T>0$ and zero temperature $T=0$ elastic moduli} \label{methodmodulus}
We computed the elastic moduli at finite temperatures $T>0$ as well as at zero temperature $T=0$ (the harmonic limit), the results of which are presented in Sec.~\ref{resultmodulus}.
In order to measure the elastic moduli, we did \textit{not} apply any explicit deformation, but rather we employed the formulation developed from the linear response theory.
For calculation of the finite $T>0$ moduli, we used the equilibrium fluctuation formulae~\cite{Lutsko_1988,Yoshimoto_2004,Mayr_2009,Wittmer_2013}, which we referred to as ``fully local approach" in our previous study~\cite{Mizuno_2013}.
For calculation of the zero $T=0$ moduli, we extended the formulation for the global moduli, which has been established and used in previous works~\cite{Lutsko_1989,Lemaitre_2006,Zaccone_2011}, to the local moduli.
In our recent work~\cite{Mizuno_2016}, we have employed this extended formulation to study local elastic moduli distributions in $T=0$ athermal jammed solids.

The present 3-dimensional system was subdivided into a grid  $40\times 40\times 40$ cubic domains of linear size $w=3.16$ (coarse graining length).
For each cubic domain $m$ ($m=1,2,...,64000$), we measured the local modulus tensor, $C^m_{\alpha \beta \gamma \delta}$ ($\alpha,\beta,\gamma,\delta = x,y,z$), which is defined as the derivative of the local stress tensor with respect to the linear strain tensor~\cite{Mizuno_2013}.
The value of $C^m_{\alpha \beta \gamma \delta}$ is composed of four terms; the Born term $C_{\alpha \beta \gamma \delta}^{Bm}$, the kinetic term $C_{\alpha \beta \gamma \delta}^{Km}$, the stress correction term $C_{\alpha \beta \gamma \delta}^{Cm}$, and the non-affine term $C_{\alpha \beta \gamma \delta}^{Nm}$;
\begin{equation}
\begin{aligned}
C_{\alpha \beta \gamma \delta}^m & = C_{\alpha \beta \gamma \delta}^{Bm} + C_{\alpha \beta \gamma \delta}^{Km} + C_{\alpha \beta \gamma \delta}^{Cm} - C_{\alpha \beta \gamma \delta}^{Nm}, \\
&= C_{\alpha \beta \gamma \delta}^{Am} - C_{\alpha \beta \gamma \delta}^{Nm}.
\end{aligned} \label{moduluseq}
\end{equation}
The summation of the first three terms, i.e. $C_{\alpha \beta \gamma \delta}^{Am}=C_{\alpha \beta \gamma \delta}^{Bm} + C_{\alpha \beta \gamma \delta}^{Km} + C_{\alpha \beta \gamma \delta}^{Cm}$, is the ``affine" modulus~\cite{Wittmer_2013,Mizuno2_2013,Mizuno_2014}.
Here we note that the Born term $C_{\alpha \beta \gamma \delta}^{Bm}$ is defined as the second derivative of the energy density with respect to the Green-Lagrangian strain tensor~\cite{Lutsko_1988,Lutsko_1989}.
Therefore, if we define the modulus by using the linear strain tensor, the stress correction term $C_{\alpha \beta \gamma \delta}^{Cm}$ is necessary as long as the stress tensor has finite values in its components~\cite{Barron_1965}.
The affine modulus describes the elastic response, when particles follow the applied affine strain field and are displaced affinely at all scales.
On the other hand, the forth term, $C_{\alpha \beta \gamma \delta}^{Nm}$, is referred to as the ``non-affine" modulus, which contributes negatively to the overall modulus, and comes from ``additional" particle displacements at the microscopic scale that deviate from the applied affine field~\cite{Wittmer_2002,Tanguy_2002}.
Finally, from the modulus tensor $C^m_{\alpha \beta \gamma \delta}$, we calculated the bulk modulus $K^m$, pure shear modulus $G^m_p$, and simple shear modulus $G^m_s$ in the same way as in Refs.~\cite{Mizuno_2013,Mizuno2_2013,Mizuno_2014}.

After measuring the local elastic moduli in each little cube $m$ ($m=1,2,...,64000$), we collected data to obtain probability distribution functions $P(M^m)$ for the moduli, $M^m=K^m, G_p^m, G_s^m$.
From $P(M^m)$, we calculated the average (global) value $M$ and the standard deviation (fluctuations) $\delta M$;
\begin{equation}
\begin{aligned}
M &= \int M^m P(M^m) dM^m, \\
\delta M &= \sqrt{\int (M^m-M)^2 P(M^m) dM^m}.
\end{aligned} \label{sds}
\end{equation}
Two shear moduli distributions, $P(G^m_p)$ and $P(G^m_s)$, coincide with each other in the isotropic systems like glasses, so we represent $G^m$ for the shear moduli in glasses, i.e. $P(G^m) \equiv P(G^m_p) \equiv P(G^m_s)$.
For reference, Table~\ref{tmoduli} presents values of the global moduli, $M=K, G_p, G_s$, for glass $2$ and crystal $2$ at zero temperature, $T=0$, and a finite temperature, $T=10^{-3}$.

In the present study, our major focus is on non-affine contributions $C_{\alpha \beta \gamma \delta}^{Nm}$, since anharmonic effects are directly reflected in $C_{\alpha \beta \gamma \delta}^{Nm}$, while affine components $C_{\alpha \beta \gamma \delta}^{Am}$ are mainly determined by static structural properties and are therefore rather insensitive to anharmonicities.
At finite $T>0$, $C_{\alpha \beta \gamma \delta}^{Nm}$ is measured in terms of a local and global stress correlation function~\cite{Lutsko_1988,Yoshimoto_2004,Mayr_2009,Wittmer_2013};
\begin{equation}
\begin{aligned}
C_{\alpha \beta \gamma \delta}^{Nm} & = \frac{V}{T} \left[ \left< \sigma^m_{\alpha \beta} \sigma_{\gamma \delta} \right>-\left< \sigma^m_{\alpha \beta} \right>\left< \sigma_{\gamma \delta} \right> \right], \\
&= \frac{V}{T} \left< \delta \sigma^m_{\alpha \beta} \delta \sigma_{\gamma \delta} \right>,
\end{aligned} \label{moduluseq2}
\end{equation}
where $\sigma_{\alpha \beta}$ and $\sigma^m_{\alpha \beta}$ are the global and local stress tensors, respectively (see Ref.~\cite{Mizuno_2013} for a detailed formulation), and $\delta \sigma_{\alpha \beta} = \sigma_{\alpha \beta}- \left< \sigma_{\alpha \beta} \right>$ and $\delta \sigma^m_{\alpha \beta} = \sigma^m_{\alpha \beta}- \left< \sigma^m_{\alpha \beta} \right>$ are the corresponding fluctuations.
$\left< \right>$ denotes the $NVT$ ensemble average, which coincides with the average over the MD trajectory when the dynamics is ergodic~\cite{Allen1986}.
Since the function of $C_{\alpha \beta \gamma \delta}^{Nm}$ contains three- and four-point correlations~\cite{Xu_2012,Wittmer_2013}, we therefore need a long simulation trajectory for good numerical convergence of the ensemble average~\cite{Mizuno_2013}.
To that end, we performed production runs for $2\times 10^{7}$ to $4\times 10^{7}$ simulation steps at each temperature $T$, and the averaging was performed over $10^4$ to $2\times 10^4$ configurations separated by $2 \times 10^3$ steps for glasses, and over $10^5$ to $2\times 10^5$ configurations separated by $2 \times 10^2$ steps for crystals.

In the $T=0$, harmonic limit, the non-affine term $C_{\alpha \beta \gamma \delta}^{Nm}$ is formulated in terms of the eigen frequency $\omega^k$ and the eigen vector $\vec{e}^k$~\cite{Lutsko_1989,Lemaitre_2006,Zaccone_2011};
\begin{equation}
\begin{aligned}
C_{\alpha \beta \gamma \delta}^{Nm} & = \sum_{k=1}^{3N-3} \frac{V}{{\omega^k}^2} \left(\sum_{i=1}^{N} \vec{e}^k_i \cdot \deri{\sigma^{m}_{\alpha \beta}}{\vec{r}_i} \right) \left(\sum_{j=1}^{N} \vec{e}^k_j \cdot \deri{\sigma_{\gamma \delta}}{\vec{r}_j} \right),\\
& = \sum_{k=1}^{3N-3} \frac{V}{{\omega^k}^2} \delta \sigma^{mk}_{\alpha \beta} \delta \sigma^k_{\gamma \delta},
\end{aligned} \label{moduluseq3}
\end{equation}
where $\delta \sigma^k_{\alpha \beta}=\sum_{i=1}^{N} \vec{e}^k_i \cdot {\partial \sigma_{\alpha \beta}}/{\partial \vec{r}_i}$ and $\delta \sigma^{mk}_{\alpha \beta}=\sum_{i=1}^{N} \vec{e}^k_i \cdot {\partial \sigma^{m}_{\alpha \beta}}/{\partial \vec{r}_i}$ are the fluctuations of the global and local stress tensors, induced by the eigen mode $k$.
We note that Eq.~(\ref{moduluseq3}) is obtained by taking the limit of $T\rightarrow 0$ in the finite $T>0$ formulation, Eq.~(\ref{moduluseq2})~\cite{Lutsko_1989}.

\bibliographystyle{apsrev4-1}
\bibliography{reference}

\end{document}